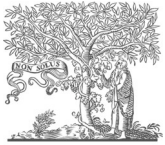
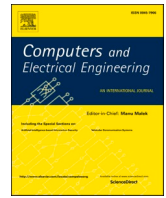

# Design and implementation of a real-time onboard system for a stratospheric balloon mission using commercial off-the-self components and a model-based approach

Ángel-Grover Pérez-Muñoz[a], Jose-Carlos Gamazo-Real[b,c,*], David González-Bárcena[c], Juan Zamorano[a]

[a] *Universidad Politécnica de Madrid. Department of Architecture and Technology of Information Systems, ETSI Ingenieros Informáticos, 28660, Madrid, Spain*
[b] *Universidad Politécnica de Madrid. Information Systems Department, Computer Architecture and Technology Area, ETSI Sistemas Informáticos, 28031, Madrid, Spain*
[c] *Universidad Politécnica de Madrid. Instituto Universitario de Microgravedad "Ignacio Da Riva", ETSI Aeronáutica y del Espacio, 28040, Madrid, Spain*

ABSTRACT

Stratospheric balloons have emerged as an affordable and flexible alternative to traditional spacecrafts as they are implemented using commercial off-the-shelf (COTS) equipment without following strict methodologies. HERCCULES is a stratospheric balloon mission that aims to characterize the convective heat and radiative environment in the stratosphere. The purpose of this article is to present the HERCCULES onboard software (OBSW) whose design and complexity is comparable to that of satellite systems, since it must control about sixty COTS equipment using a single Raspberry Pi 4B as onboard computer and ensure the real-time requirements. Compared to similar systems, novel contributions are presented as the OBSW is developed following model-based and component-based approaches using the TASTE toolchain from the European Space

*Abbreviations:* AADL, Architecture Analysis and Design Language; ASN.1, Abstract Syntax Notation 1; ASSERT, Automated proof-based System and Software Engineering for Real-Time applications; ATL, Attitude Laboratory; BEXUS, Balloon EXperiments for University Students; BSP, Board Support Package; CBD, Component-Based Development; CDPI, Combined Data and Power management Infrastructure; CONOPS, CONcept of OPerationS; COTS, Commercial Off-The-Shelf; CV, Concurrency View; DEL, Downwards Environmental Laboratory; DLR, German Aerospace Center; DPV, Deployment View; DV, Data View; E-Box, Electronic Box; EL, Environmental Laboratory; ELS, Environmental Laboratory Support; ESA, European Space Agency; FIFO, Firs-In First-Out; GPS, Global Positioning System; GS, Ground Station; HAL, Hardware Abstraction Layer; HERCCULES, Heat-transfer and Environment Radiative and Convective Characterization in a University Laboratory for Experimentation in the Stratosphere; HTL, Heat Transfer Laboratory; HK, Housekeeping; $I^2C$, Inter-Integrated Circuit; IDR, Instituto de Microgravedad Ignacio Da Riva; IMU, Inertial Measurement Unit; IV, Interface View; MBD, Model Based Development; MIAT, Minimum Inter-Arrival Time; MSE, Mean Squared Error; NADS, Navigation and Attitude Determination Subsystem; OBC, On-Board Computer; OBDH, On-Board Data Handling; OBSW, On-Board Software; OS, Operating System; OSAL, Operating System Abstraction Layer; PCB, Printed Circuit Board; PCU, Power Control Unit; PI, Provided Interface; PIM, Platform Independent Models; PSM, Platform Specific Models; PWM, Pulse Width Modulation; RI, Required Interface; RTA, Response Time Analysis; RTES, Real-Time and Embedded System; SC, Scientific; SDPU, Sensor Data Processing Unit; SNSA, Swedish National Space Agency; TASTE, The ASSERT Set of Tools for Engineering; TC, Telecommand; TM, Telemetry; TMU, Thermal Measurement Unit; UEL, Upwards Environmental Laboratory; UPM, Universidad Politécnica de Madrid; UART, Universal Asynchronous Receiver-Transmitter; V&V, Verification and Validation; WCET, Worst Case Execution Time.

* Corresponding author.
  *E-mail addresses:* josecarlos.gamazo@upm.es, jcgamazo@gmail.com (J.-C. Gamazo-Real).

Agency (ESA) for automatic code generation. Besides, the OBSW is verified and validated following the ESA standards and the results obtained demonstrate the suitability and efficiency of the solution and the selected methodologies.

## 1. Introduction

The number of stratospheric flights in the last years increased considerably not only due to their ease of operation and reduced cost compared to other space vehicles [1], but also because of their wide range of applications such as Earth observation, atmosphere characterization, telecommunications, space exploration, technology demonstration, etc. [2]. The mission payloads, which include the required systems to reach the mission goals, are attached to a balloon inflated by helium that rises to the stratosphere (around 30 km) remaining there for a few hours up to few weeks. Once the mission is finished, the balloon is prickled, and the payload can be recovered, providing some advantages compared to traditional spaceflights such payload reusage and recovery of data recorded onboard the system. In this manner, the requirements for onboard equipment are less critical, opening the possibility of using commercial off-the-shelf (COTS) components that are not traditionally applied in safety-critical systems.

The environment in the stratosphere is similar to space since it has near-vacuum conditions and relies on radiation as the main mechanism of heat transfer with the environment [3]. Nevertheless, the environmental conditions are usually more stable than in orbit, where the illuminated and eclipse cycles could drive the system to extreme temperature conditions. Compared to Earth's surface conditions, in the stratosphere, the elevated temperatures caused by the absence of convection can damage electronics degrading their performance. Typically, this problem is solved by cooling the components, but in not pressurized stratospheric experiments other solutions should be considered, such as increasing the conductive coupling to a cooler interface, maximizing radiation, or distributing the heat in the printed circuit board (PCB) [4]. The use of COTS components seems feasible in stratospheric experiments in which the air pressure is not high enough for outgassing, the reliability of the technology is not as critical as in satellites, and the mass and power requirements are usually less restricted. From a software perspective, the main advantage of COTS is reusability, as the software driving COTS components is often provided by the manufacturer or available in the open-source community. This reduces the time to release, and the development is focused on the implementation of requirements rather than coding or maintenance.

This article presents the soft real-time onboard software (OBSW) of the HERCCULES (Heat-transfer and Environment Radiative and Convective Characterization in a University Laboratory for Experimentation in the Stratosphere) mission. The main objective of HERCCULES is to characterize the convective and radiative environment in the stratosphere to improve the thermal modeling of stratospheric systems. As a secondary objective, the performance of a Nadir sensor as an attitude determination instrument is evaluated. HERCCULES is design with COTS technology, and to comply with its objectives, it implements several experiments such as the Heat Transfer Laboratory (HTL) equipped with heated plates and thermistors, or the Environmental Laboratory (EL) composed of radiometers to monitor radiation from different fields of view. HERCCULES was selected by the European Space Agency (ESA) for the Balloon EXperiments for University Students (BEXUS) program which is realized under a bilateral agreement between the German Aerospace Center (DLR) and the Swedish National Space Agency (SNSA). Besides, HERCCULES is supported by the "Instituto Universitario de Microgravedad Ignacio Da Riva" (IDR) and the "Sistemas de Tiempo Real y Arquitectura de Servicios Telemáticos" (STRAST) research group from the Universidad Politécnica de Madrid (UPM). The launch of the mission is scheduled for 25th September 2023 from the SNSA launch base in Esrange (Kiruna, Sweden) and the flight will last about 6 hours.

The main objective of this article is to present the methodology and design solutions of the OBSW for the HERCCULES stratospheric balloon. The system consists of a central onboard computer (OBC) which is a Raspberry Pi 4B board running a Linux operating system (OS). The OBSW deals with the complexity of the HERCCULES system that controls the near sixty peripherals, including radiometers, thermometers, barometers, photodiodes, and heaters, among others. The OBSW performs the Onboard Data Handling (OBDH) activities including the storage of scientific (SC) and housekeeping (HK) telemetries (TM) for post-mission analysis and their periodic transmission to a remote Ground Station (GS). The OBSW adopts Model-Based Development (MBD) and Component-Based Development (CBD) approaches. Specifically, the OBSW is developed with The ASSERT Set of Tools for Engineering (TASTE) [5] from ESA that targets real-time and embedded systems (RTES). The presented OBSW solution addressed successfully all these challenges addressing all real-time requirements and applying all the aforementioned tools and methodologies.

It is noteworthy that there is significant research on ballooning and CubeSats missions based on COTS components such as [6–8], but they do not provide details about the real-time design, software development, verification, or validation methodologies adopted in their OBSW, nor numerical assessment of the performance and accuracy of their measurement results. As a secondary objective, this article aims to address this gap by discussing the processes and design patterns applied in the HERCCULES OBSW. This article presents three design patterns for RTES that were successfully applied in the OBSW design. In addition, the software is verified and validated following an incremental approach at the unitary, integration and system levels. For this purpose, four methods based on the ECSS-E-ST-10-02C standard [9] are adopted namely Test, Analysis, Design Review, and Inspection. Automated Tests are used to compare the obtained measurements with reference values. Inspection and Analysis are applied in the absence of reference value. The Analysis is achieved calculating the Mean Squared Error (MSE) and percentage error. These statistical metrics are typically used in artificial intelligence systems such as in [10] that uses the MSE to assess the performance of a machine learning model used to reduce energy consumption. Finally, the source code quality is analyzed by comparison with values recommended for safety-critical space systems (level A). The obtained results not only demonstrate their suitability for this specific mission but also suggest their applicability in more general control and data acquisition systems like ground-based weather stations.

Based on these considerations, the main contributions of this article can be summarized as follows:

- A software solution is proposed for the HERCCULES stratospheric balloon. The software architecture and design can be extrapolated to systems with similar complexity as weather stations or OBDH satellite systems.
- The HERCCULES OBSW implementation is based on design patterns suitable for RTES. Such patterns are described in detail in this paper and can serve as a reference for a wide range of applications.
- The studies carried out demonstrate the value and benefits of the CBD and MBD methodologies, both are effectively applied for the HERCCULES OBSW development.
- Finally, although the V-model and ECSS standards are commonly reserved for complex space projects, it is adopted in HERCCULES, a non-critical system based on COTS, improving the overall reliability of the system.

The remainder of this article is structured as follows. Section 2 presents a general overview of the HERCCULES mission and subdivision in experiments and subsystems. The Section 3 describes the life cycle (V-model), research and development methodologies (MBD and CBD), and technologies (TASTE) adopted for the OBSW development. The Section 4 deals with the OBDH system architecture, focusing on the OBC and the communication with the experiments. Section 5 presents the OBSW architecture and design patterns used to implement it. The experimental results, comparison to related works, and future works are discussed in Section 6. Finally, the conclusions are drawn in Section 7.

## 2. HERCCULES mission overview

### 2.1. Mission operations

The concept of operations (CONOPS) describes the system from the operators' points of view and helps to define the core OBSW functionality since it covers the operations and events triggered during the mission. The operational modes include different states that define the activities performed by each subsystem while in the state. Fig. 1 illustrates the five HERCCULES phases. The mission begins in a *Pre-Launch* minutes before the *Launch* with all experiments powered off. In these phases, the operators turn on the experiments to test the equipment functionality and ensure communication with the GS. When the balloon inflation is completed the *Ascent* phase starts and all subsystems are powered on by the OBSW to start the data acquisition. The *Float* phase is entered at 25-30 km (stratosphere) above sea level, approximately after 1.5 hours from Launch. All subsystems remain unchanged, except for the HTL that controls heaters with different power dissipation. The *Descent* phase is reached after approximately five hours when the cutter separates the balloon from the gondola. In this phase, subsystems and heaters are turned off and the OBSW stops. Finally, after the *Recovery* phase, where the experiment is transported back to the launch base, the acquired data is analyzed, and the results are reported. It is worth noting that Ascent and Floating correspond to nominal phases, as in those mode the HK and SC data is collected and sent to the GS at 1 and 10 second periods, respectively. Besides, HERCCULES includes a manual mode to control the system by means of telecommands (TC) and an autonomous mode where automatic control is performed.

### 2.2. Experimental setup

HERCCULES was divided into five subsystems that were designed using COTS modules to measure thermal data such as radiation, temperature, pressure, and data for attitude determination. Subsystems were equipped with actuators to control heated plates and payload instrument temperatures. The context diagram of the system is depicted in Fig. 2-a. It should be noted that the E-Link system was not part of HERCCULES; rather, it was provided by the BEXUS-32 program to communicate GS and OBC. Relevant characteristics of E-Link included Ethernet 10/100 Base-T protocol, S-band operating frequency, 2 Mbps duplex nominal bandwidth shared across experiments, and conventional RJ45 interfaces [11].

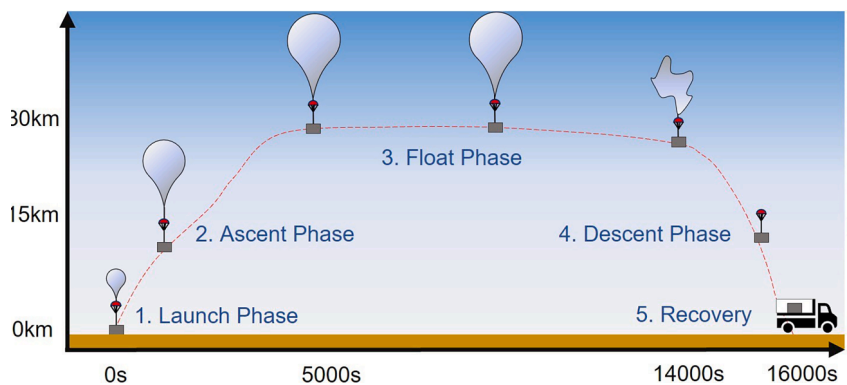

**Fig. 1.** Definition of the HERCCULES mission CONOPS.

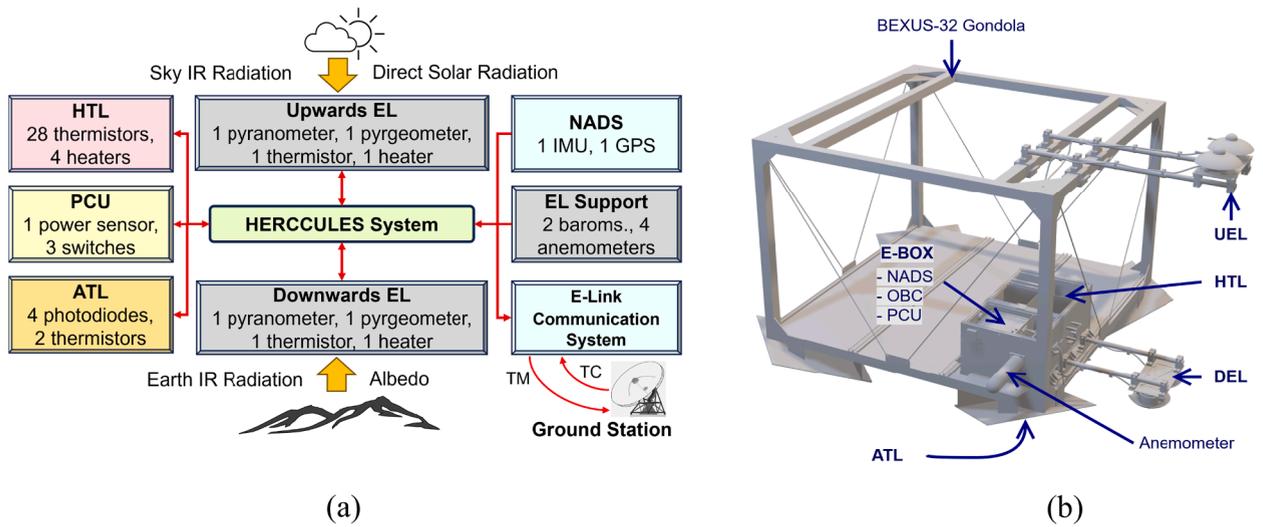

**Fig. 2.** Experimental setup of HERCCULES: context diagram (a) and position inside the gondola (b).

Fig. 2-b presents the general structure of the BEXUS-32 gondola which has dimensions of 1.16 m x 1.16 m x 0.84 m. The HERCCULES system shares this gondola with other university experiments, and its subsystems are placed in different configurations. The first subsystem is the HTL that quantified the heat transfer through air. It was divided into four experiments composed of several aluminum plates whose temperature was monitored by twenty-eight PT1000 thermistors and controlled through four silicon heaters. PT1000 thermistors were selected as they could be mounted directly on the measuring body with good accuracy. Secondly, the EL characterized the thermal environment by measuring radiation with pyranometers and pyrgeometers, air pressure with two barometers, and relative wind speed with four differential barometers. The radiometers' temperature was controlled with PT1000 thermistors and silicon heaters. The EL was arranged in three compartments: the Environmental Lab Support (ELS) was placed inside the gondola, while the Upwards and Downwards Environmental Laboratory (UEL and DEL) were located outside. Thirdly, the Attitude Laboratory (ATL) consisted of a Nadir sensor based on four photodiodes that measured the infrared radiation. To validate ATL, the Navigation and Attitude Determination Subsystem (NADS) collected attitude data from one Global Positioning System (GPS) receiver and one Inertial Measurement Unit (IMU) providing the linear acceleration, magnetic field strength, and angular velocity. Finally, the PCU distributed the incoming 28.8 V and 1 mA from a battery pack at three voltage levels: 12V, 5V, and 3.3 V. The PCU included three switches to control power distribution and voltage/current sensors that served as HK TM data.

## 3. Research and working methodology

### 3.1. Research methodology

The presented work follows a research methodology consisting of three phases as depicted in Fig. 3. The first phase refers to the literature review where it was analyzed the state of the art on technologies, methodologies, and architectural and design patterns suitable for RTES, especially those applied in the space sector. This phase also involved the analysis of similar works such as micro-satellites, CubeSats, weather stations and ballooning applications. At the end of the literature review, the development phase started and the objectives, CONOPS and requirements of the mission were obtained. Subsequently, the design phase started, which consisted of the elaboration of the functional and real-time architecture of the OBSW. These two activities were carried out with TASTE following the MBD and CBD approaches. Moreover, the design patterns discovered in the literature review were effectively applied in the OBSW design. Subsequently, the implementation and the verification and validation (V&V) of the system were conducted. Finally, the results obtained from the V&V phases were analyzed and allowed a quantitative and qualitative comparison with related research works. This comparison and the results enabled to draw the conclusions and challenges that would be addressed in future works.

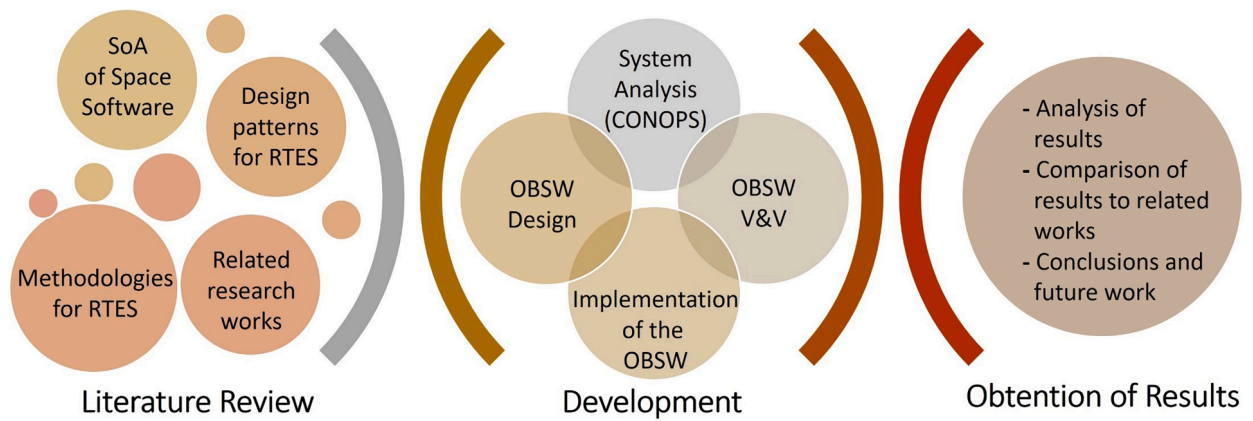

**Fig. 3.** Research methodology applied to conduct the study.

*3.2. Software life cycle*

The HERCCULES OBSW fell into the category of soft RTES since it interacted with the physical environment through a set of equipment that responded to input stimuli within a finite and specified period [12]. It is said to be "soft" because occasional deadline misses can be ignored and services can be occasionally delivered late with an upper limit of tardiness [12]. The software life cycle required the early discovery of errors, especially those related to synchronization and functional correctness. The applied methodology was a V-model which is usually reserved for safety-critical systems due to its complexity and the necessity of well-defined activities. Although this may be perceived as a disadvantage, this also makes it suitable for space systems as the requirements are clearly defined earlier. Besides, the V-model provides relevant advantages because it highlights the importance of the software V&V activities by defining specific test and review plans per development phase. Fig. 4 illustrates the V-model workflow and its relationship with the review phases of HERCCULES. In this graph, the left side of the model shows the development activities, where the requirements and specifications are defined. The right side refers to the V&V performed on the outputs of the development phase.

*3.3. Software development process*

The OBSW was developed following the MBD and CBD paradigms. In the realm of safety-critical space systems, MBD is of special interest because it allows engineers to validate attitude control systems using the "in-the-loop" techniques simulating the environment characteristics such as radiation, gravity, or acceleration [13]. In addition, MBD provides automatic code generation capabilities reducing time and effort and avoiding error-prone tasks by ensuring system correctness. The OBSW was developed with the ESA's TASTE open-source toolchain [5]. TASTE is the result of the Automated proof-based System and Software Engineering for Real-Time applications (ASSERT) project and follows the MBD and CBD paradigms to support the ASSERT process (Fig. 5-a) and transform models into source code (Fig. 5-b).

Depending on their level of platform dependency, the models were divided into different categories and refined until the source code was generated. These models were represented by TASTE in the Architecture with the Analysis and Design Language (AADL). The first type refers to the *Platform Independent Models* (PIM) that capture the functional behavior of the system independently from the execution platform. PIM includes the *Data View* (DV) models that define the data types in Abstract Syntax Notation One (ASN.1) and *Interface View* (IV) models that capture the functional elements and their relationships. The second type are the *Platform Specific Models* (PSM), which includes the *Deployment View* (DPV) models that specify the physical platforms in which the functional elements from the IV are deployed, such as processors and buses. PSM contain the real-time and concurrency properties of the system (task release time, stack size, etc.) in the *Concurrency View* (CV) models. PIM and PSM models are automatically transformed into source code. TASTE uses the ASN1SCC compiler to transform the data types specified in the DV (ASN.1) into C and Ada code, while XML files are translated to AADL with the XML2AADL tool. Finally, the Kazoo tool makes use of Ocarina to generate the PolyORB-Hi middleware that abstracted general services such as file management or task creation.

The real-time requirements were reflected in these models thanks to the support of TASTE for real-time abstractions. Specifically, TASTE Functions in the IV offer operations through Provided Interfaces (PI) and access others through Required Interfaces (RI). The PIs are classified as synchronous and asynchronous. Firstly, synchronous PIs execute on the context of the caller task and they may be either unprotected or protected to ensure mutual exclusion. Asynchronous PIs execute on a dedicated task and they may be either event-based (sporadic) or time-based activated (cyclic). Besides, TASTE allows developers to assign properties such as minimum inter-arrival time (MIAT) for sporadic PIs, worst case execution time (WCET) for all types of PIs, deadlines for asynchronous PIs, and priorities for asynchronous and protected PIs. These are represented as AADL properties so that schedulability and response-time analysis (RTA) can be performed.

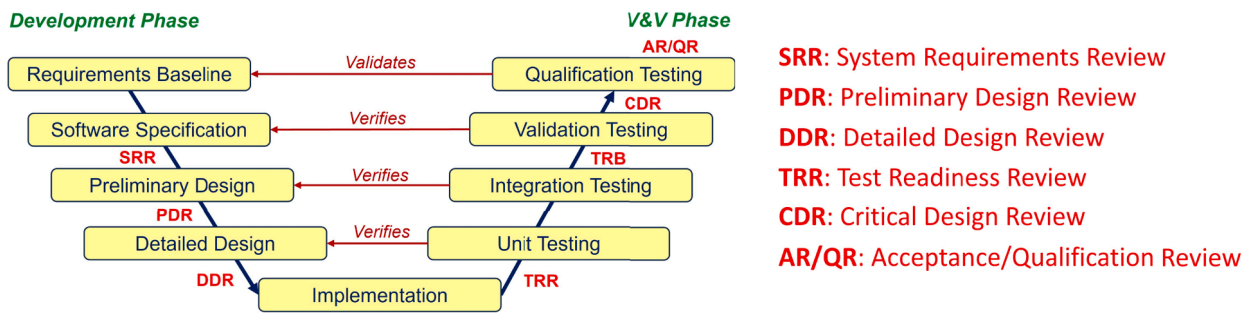

**Fig. 4.** V-model lifecycle for space domain software applied in the HERCCULES mission.

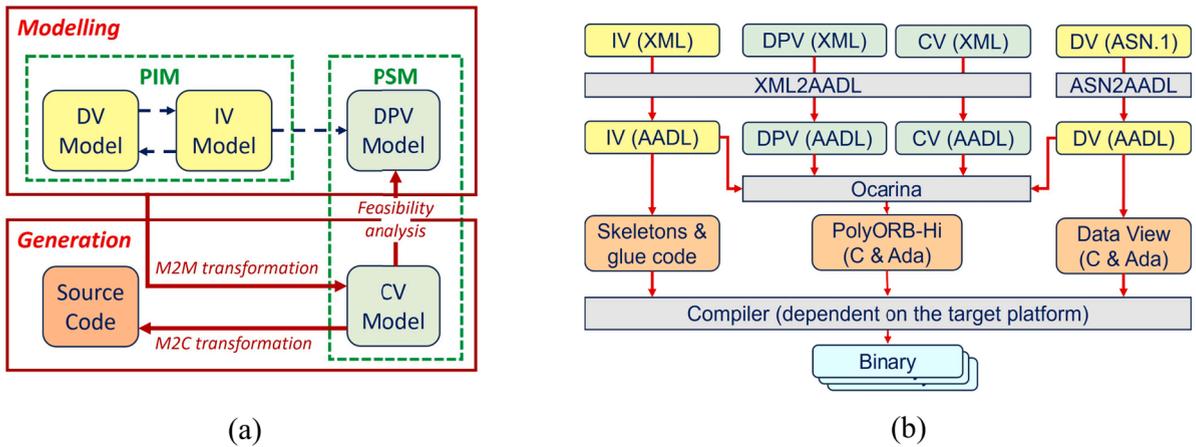

**Fig. 5.** ASSERT model-based development process (a) and code generation process of the TASTE toolchain (b). Color coding: yellow for the PIM models, green for the PSM models, and orange for the autogenerated code.

## 4. Onboard data handling

OBDH systems are the central component of satellite missions, as they are responsible for the communication between the functional units of the spacecraft and the GS. They implement the execution of TCs, generation of TM, health status monitoring, and failure detection and isolation recovery (FDIR), among other activities. The HERCCULES OBDH was based on a central OBC architecture that controlled, supervised, and acquired data from the system equipment through remote terminal units (RTU). RTUs were external I/O boards connected to the platform equipment and payload instruments by digital lines, such as Inter-Integrated Circuit ($I^2C$) or Universal Asynchronous Receiver Transmitter (UART), and analog lines through analog-to-digital converters (ADC) and multiplexers. Eickhoff and Airbus DS [14] conceptualized this as the Combined Data and Power Infrastructure (CDPI), which was successfully adopted in previous missions such as UPMSat-2 [15] and Flying Laptop Generation (FLP) 1 and 2 [14]. Fig. 6-a shows the generic topology of this architecture in which the power management system acted as a special RTU distributing power across systems. The RTUs were connected to the OBC inside the electronic box (E-Box), as shown in Fig. 6-b, and interfaced with the five subsystems to cover the analog-to-digital (A/D) conversion, power, and I/O distribution. The *Onboard Computer* corresponded to the Raspberry Pi, the *Thermal Measurement Unit* (TMU) and *Sensor Data Processing Unit* (SDPU) provided signal conditioning and multiplexing for the PT1000 thermistors through one ADC and four 8:1 multiplexers, and the *Power Control Unit* (PCU) provided power distribution across all other RTUs and pulse-width modulation (PWM) control of heaters.

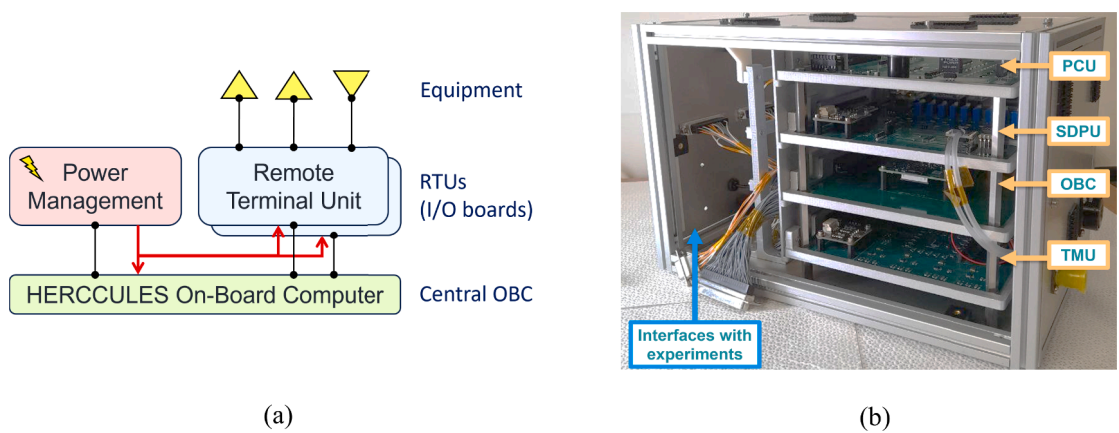

**Fig. 6.** CDPI generic architecture (a) and its application in the HERCCULES mission (b).

## 4.1. Onboard computer

OBCs for space missions require radiation-hardened and fault-tolerant processors since they are subject to harsh environmental conditions, such as temperature, radiation, pressure, vibration, and acceleration. This does not apply to HERCCULES as the system will fly in the stratosphere at approximately 30 km. The OBC selected for HERCCULES was the COTS-type Raspberry Pi Model 4B. This single-board computer was successfully used in previous BEXUS and ballooning applications due to its powerful hardware features and availability of software development tools. The Raspberry Pi 4B contains a Broadcom BCM2711 system-on-a-chip with a 64-bit quad-core ARM Cortex-A72 microprocessor, implementing the ARMv8 instruction set and running at 1.5 GHz. The processor board also includes 8 GB of LPDDR4 RAM for running the OBSW, 64 GB of flash memory to deploy the OBSW and store TM data, 1 Gigabit Ethernet port for communications with the GS, and 40 general-purpose input/output (GPIO) lines that support up to six UART ports, six $I^2C$ ports, five Serial Peripheral Interfaces (SPI) and two PWM interfaces.

The Raspberry Pi OS was used in this mission since it is the official Debian-based OS for Raspberry Pi SBCs. Although it is not a real-time OS (RTOS), it is an embedded Linux OS. As such, it is compliant with the Portable Operating System Interface X (POSIX), providing operations for thread management and the fixed priority real-time scheduling, namely SCHED_FIFO. Most critical applications would require the PREEMPT_RT patches to use Raspberry Pi OS as an RTOS; that is not the case in HERCCULES which used the default configuration. In addition to the hardware-supported protocols like UART, $I^2C$, and SPI, Raspberry Pi OS offers built-in bit-bang variants to replace such controllers with software. This is of special interest when the number of hardware interfaces is not sufficient, as in the HERCCULES mission.

## 4.2. Communication interfaces

The onboard equipment was connected to the OBC as peripherals via RTUs. They had direct connections for digital sensors utilizing GPIO pins, UART or $I^2C$, and indirect connections for analog equipment connected through adaptation circuits, such as ADCs and multiplexers. Fig. 7 shows hardware interfaces and the relationship of the OBC to the external equipment through both types of connections. The bus hierarchy was designed such that analog sensors sharing a common $I^2C$ bus and ADC had similar sampling periods, and preferably belonged to the same subsystem, such as the UEL and DEL sensors connected to the I2C-3 interface. Specifically, two digital absolute barometers (ABS-BAR) were connected to this bus, and analog sensors such as the pyranometers (PYRA), pyrgeometers (PYRG), and differential barometers (DIFF-BAR) interfaced the I2C-3 bus via the ADC (ADC-SDPU). Finally, eleven GPIO pins were used to power on-off the RTUs, control the heaters, and select multiplexed lines. The usage of GPIO pins is presented in Table 1. Note that signal multiplexing (using ADC and MUX) allowed to efficiently connect various analog sensors consuming few OBC pins.

## 5. Onboard software design

### 5.1. Static architecture

The UML package diagram in Fig. 8 depicts the high-level static architecture of the OBSW, where components are represented as packages and their dependency relationships as dashed arrows. Software components were used to represent modular and reusable units as layers, experiments, systems, and subsystems. The OBSW followed a layered architecture where low-level components managed operations closer to the hardware, while higher level components implemented the functional logic of the mission. Note the dependency from the OBSW on the Raspberry Pi OS through the POSIX interface, which was used to invoke system calls at the user-space. Other libraries were also imported for lower level operations.

The *Hardware Abstraction Layer* (HAL) comprised three sublayers to access the hardware. Firstly, the *Bus Handlers* wrapped the i2c-dev, termios, and pigpio libraries to access to the $I^2C$, UART, and GPIO interfaces, respectively. These operations were controlled by kernel drivers, but they were accessed at user-space through the "/dev" interface. On the other hand, the *Equipment Handlers* abstracted the control of hardware devices, and the *Board Support* abstracted the access to the SDPU, PCU, and TMU. The *Data Storage* provided an interface to store TM data in the micro-SD card. The HAL and Data Storage were passive components and did not contain any task or protected object. HERCULES followed a data-centric architecture as the *Data Pool* component gathered the data shared across tasks guaranteeing safe concurrent access through "protected objects". This concept comes from the Ada concurrency model, and it was implemented by TASTE with mutual exclusion primitives from the POSIX Thread library. The *Telemetry and Telecommand* (TTC) component implemented fault-tolerant communication with the GS and included a periodic task to send SC and HK TM from the Data Pool at 1000 ms period, and one sporadic task for redirecting incoming TCs to the Manager at 1000 ms MIAT. The *Manager* implemented system-level operations such as the management of the operating modes. It contained two sporadic tasks to handle TCs and events, both with MIATs of 1000 ms. Finally, the *Subsystems* component implemented the functionality of the subsystems (PCU, NADS, HTL, EL, and ATL) in separate cyclic tasks, with periods ranging from 10 to 10000 ms. These tasks performed the traditional data acquisition cycle: measure, control, and actuate.

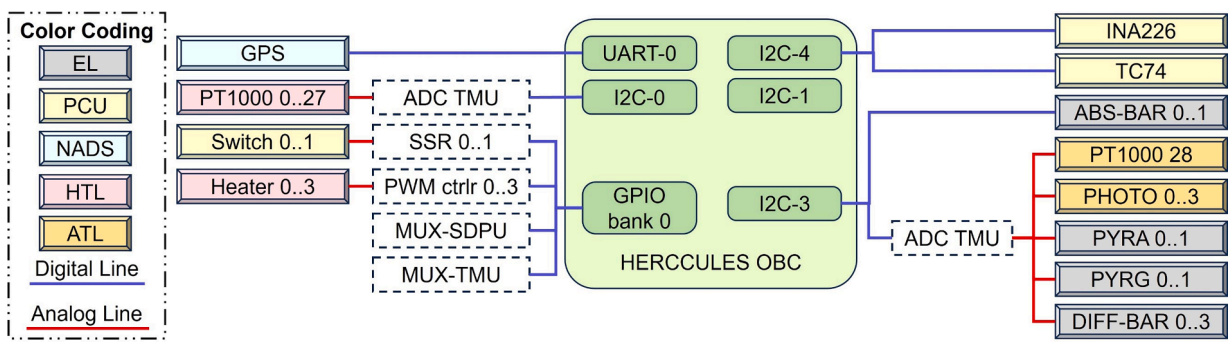

**Fig. 7.** Data communication links and hardware interfaces.

**Table 1**
Usage of GPIO pins and hardware interfaces.

| Lines | GPIO | I$^2$C | UART | PWM | Total |
| --- | --- | --- | --- | --- | --- |
| Interface use | 11 | 4 | 1 | 4 | 20 |
| GPIO pin use | 11 | 8 | 2 | 4 | 25 |

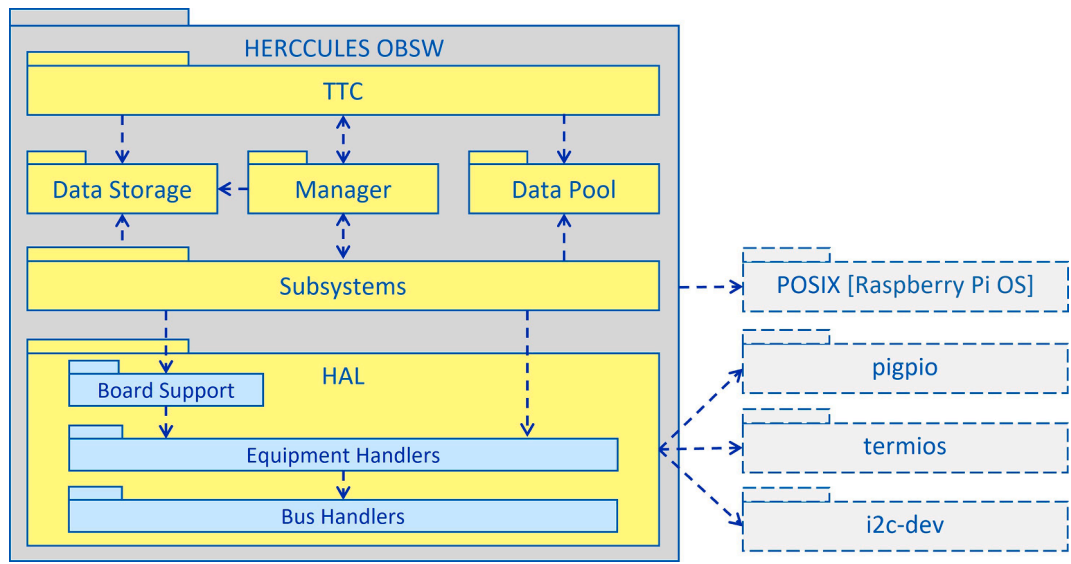

**Fig. 8.** UML package diagram from the HERCCULES OBSW.

### 5.2. Dynamic architecture

The task set from each component is listed in Table 2. The description includes activation periods or MIAT for sporadic tasks (T), deadlines (D), priorities (P), and accessed protected objects. Periods for the Subsystems tasks were assigned based on the change rates in the environment. For instance, temperature measurement from the HTL Manager required 10000 ms to perform the control cycle, but the IMU Measurer task required a period of 10 ms due to the IMU operating frequency. Similarly, the tasks Event Handler, TM Sender, TC Receiver, and TC Handler worked at a 1000 ms period according to the generation rate of events, TM, and TCs. Most of the tasks' deadline was equal to its period, except for TC Handler and Event Handler that required a rapid response to events. Tasks were assigned priorities based on the Deadline Monotonic Scheduling scheme where tasks with shorter deadlines get higher priorities. However, despite having the longest deadline, the HTL Manager task was assigned a higher priority due to its impact on the mission success. This practice aligns with the recommendation outlined in the "Guide for the use of the Ravenscar profile" [16]. Similarly, the SDPU measurer was given a higher priority than TC Receiver and TM Sender.

**Table 2**
Task set of the HERCCULES OBSW with time values (T, D, P) in milliseconds.

| Component | Task | T | D | P | Accesses to protected objects |
| --- | --- | --- | --- | --- | --- |
| Manager | TC Handler (sporadic) | 1000 | 500 | 3 | TC-Queue, NADS-Mode, NADS-Dev, HTL-Mode, HTL-Ctrlr, HTL-Dev, SDPU-Mode, SDPU-Dev, EL-Ctrlr, PCU-Mode, PCU-Dev, TTC-Mode, TTC-TM-Mode |
|  | Event Handler (sporadic) | 1000 | 500 | 3 | Event-Queue, NADS-Mode, HTL-Mode, HTL-Ctrlr, PCU-Mode, SDPU-Mode, EL-Ctrlr, TTC-Mode |
| TTC | TC Receiver | 1000 | 1000 | 2 | Event-Queue, TTC-TM-Mode, TC-Queue, DP-NADS, DP-HTL, DP-EL, DP-PCU, DP-ATL |
|  | TM Sender | 1000 | 1000 | 2 | Event-Queue, TTC-TM-Mode, TTC-Mode |
| Subsystems | IMU Measurer | 10 | 10 | 6 | NADS-Mode, DP-NADS |
|  | GPS Measurer | 200 | 200 | 5 | NADS-Mode, DP-NADS |
|  | HTL Manager | 10000 | 10000 | 4 | HTL-Mode, HTL-Ctrlr, DP-HTL, DP-EL |
|  | SDPU Measurer | 1000 | 1000 | 3 | Event-Queue, SDPU-Mode, SDPU-Ctrlr, DP-ATL, DP-EL, EL-Ctrlr |
|  | PCU Manager | 5000 | 5000 | 1 | PCU-Mode, DP-PCU |

These real-time attributes were supported by TASTE autogenerated code making use of the POSIX profile for RTES. Specifically, tasks were configured with the fixed-priority real-time scheduling policy (SCHED_FIFO) and the priority ceiling protocol for mutual exclusion objects as it prevents mutual deadlocks (PTHREAD_PRIO_PROTECT). These protocols were selected by TASTE because its runtime is based on the Ravenscar profile [16]. Although the Raspberry Pi did not include a real-time clock (RTC), absolute time was obtained from the GPS receiver and configured through the "settimeofday" system call. Regarding relative time capabilities, the default configuration of the Raspberry Pi OS kernel and hardware supported high-resolution timers (HRT), which are more accurate and suitable for RTES since they provide a resolution of nanoseconds compared to milliseconds on standard timers. The HRT support was corroborated by inspecting the resolution entries from "/proc/timer_list". This allowed TASTE to implement task delays with the CLOCK_MONOTONIC clock using the "nanosleep" system call. Finally, aiming for determinism, TASTE masked all HERCCULES OBSW tasks to run in the same processor core through the "sched_setaffinity" system call.

The UML communication diagram in Fig. 9 illustrates the dynamic behavior of the OBSW tasks. Synchronous messages (black arrows) were executed in the context of the calling task and were accessed by the receiver through a protected object that stored the latest message. On the other hand, asynchronous messages (blue arrows) represent events and TCs that were stored in the TC-Queue and Event-Queue protected first-in-first-out (FIFO) queues, which were processed by the Handler and Event Handler TCs, respectively. Two execution flows were identified. The first one referred to the change of the system mode. Whenever the Manager received an event or TC that affected the operating mode, it notified the new mode to all other systems through a synchronous message, updating the NADS-Mode, PCU-Mode, HTL-Mode, SDPU-Mode, and TTC-Mode protected objects. It is noteworthy that the EL and ATL were passive objects that received this notification from the SDPU Measurer task. The second execution flow was started by other subsystems, such as when the SDPU Measurer task detected a floating altitude or a cut-off, and then the relevant events were putted in the Event Queue. Such events activated the execution of the Event Handler sporadic task and subsequently the first scenario happened.

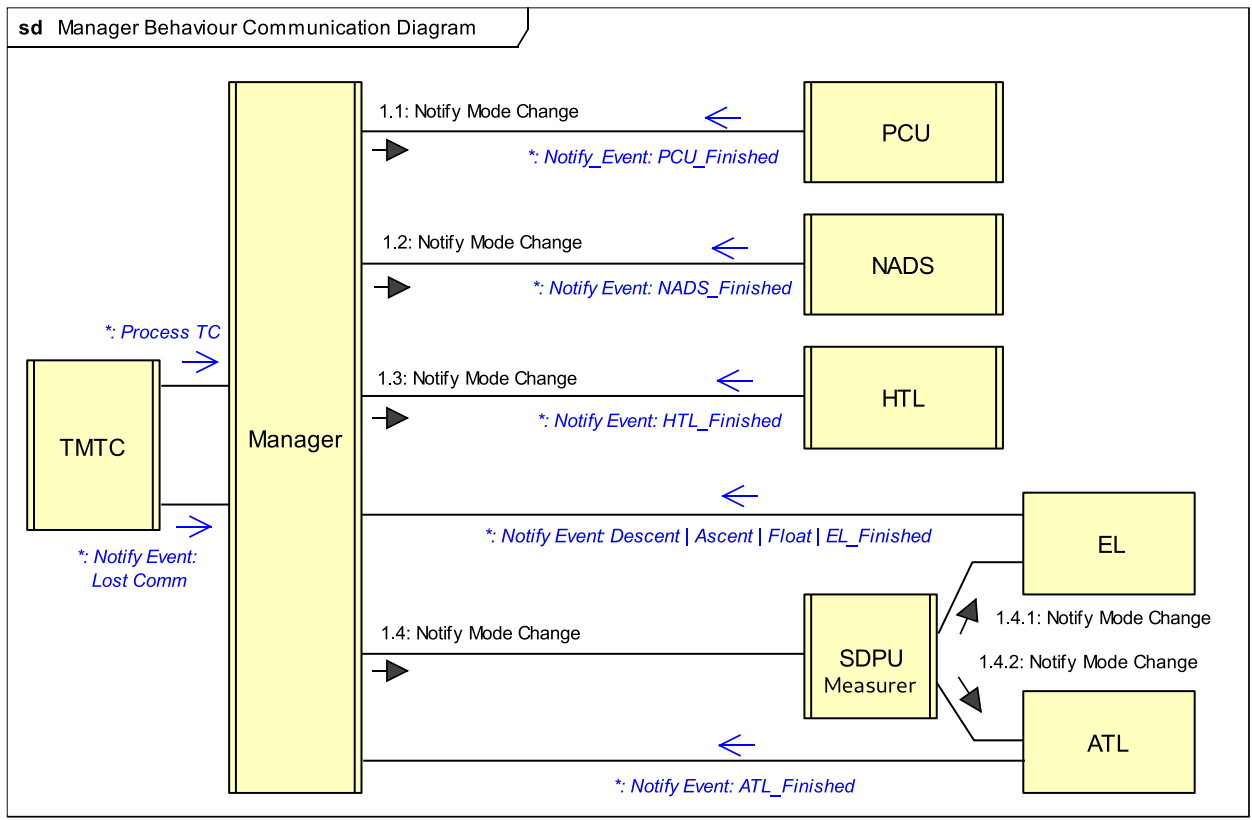

**Fig. 9.** HERCCULES OBSW dynamic architecture.

*5.3. Detailed design with TASTE*

The HERCCULES OBSW architecture was implemented with the TASTE toolchain. The relationships between components were achieved through PIs, the public operations offered to external components, and RIs, the operations required by a component to perform its operations. The IV of TASTE implementation is presented in Fig. 10. It shows the high-level components of the system represented as yellow blocks, named Functions in TASTE. This IV reflects both the static structure of the system (modeled in Fig. 8 as a package diagram) and the dynamic aspects that allow intercommunication among Functions (modeled in Fig. 9 as a communication diagram). The identified sporadic tasks were modeled as sporadic PIs, periodic tasks as periodic PIs, and protected objects as protected PIs. For instance, the Event Handler sporadic task corresponds to the "Notify Event" sporadic PI in the "HERCCULES Manager" Function. This interface was configured with a MIAT of 1000 ms and a FIFO queue to store at most ten pending events.

Regarding the dynamic architecture, the "HERCCULES Manager" contained five "Notify Mode Change" protected RIs to update the balloon mode of all subsystems. The notification sequence to handle events, errors and TCs corresponded to the event-based architecture. As depicted in the IV, this architecture was implemented following a static topology, as Functions were explicitly connected through independent PI/RI connections. This topology increased the predictability of the system, but affected scalability since every time a new subscriber was added to the system, it was necessary to modify both the interface and the implementation of the Manager, thus generating a strong coupling relationship. As for benefits, TASTE automatically generated the concurrent aspects of the system such as mutexes, tasks, delays, semaphores, and message queues, which allowed rapid system development with reduced costs and an increase in the development time.

*5.4. Architectural and design patterns for RTES*

*5.4.1. Hierarchical layered architecture*

The layered pattern organized the software into layers that offered its operations to upper layers and depended on lower layers to implement them. In the case of general RTES, this architecture is applied in many software development kits and frameworks provided by vendors to separate the core logic and the hardware-related operations. The Fig. 11 showcases two instances of this pattern. Firstly, the Fig. 11-a depicts the classical layered architecture where the drivers and board support package (BSP) provide operations to initialize, configure, and control the hardware. The HAL acts as an interface declaring these operations without revealing implementation details. The RTOS or microkernel is isolated from the application through the OS abstraction layer (OSAL). Fig. 11-b presents the structure of the code generated by TASTE, where the Application Layer executed at user-space and was manually coded.

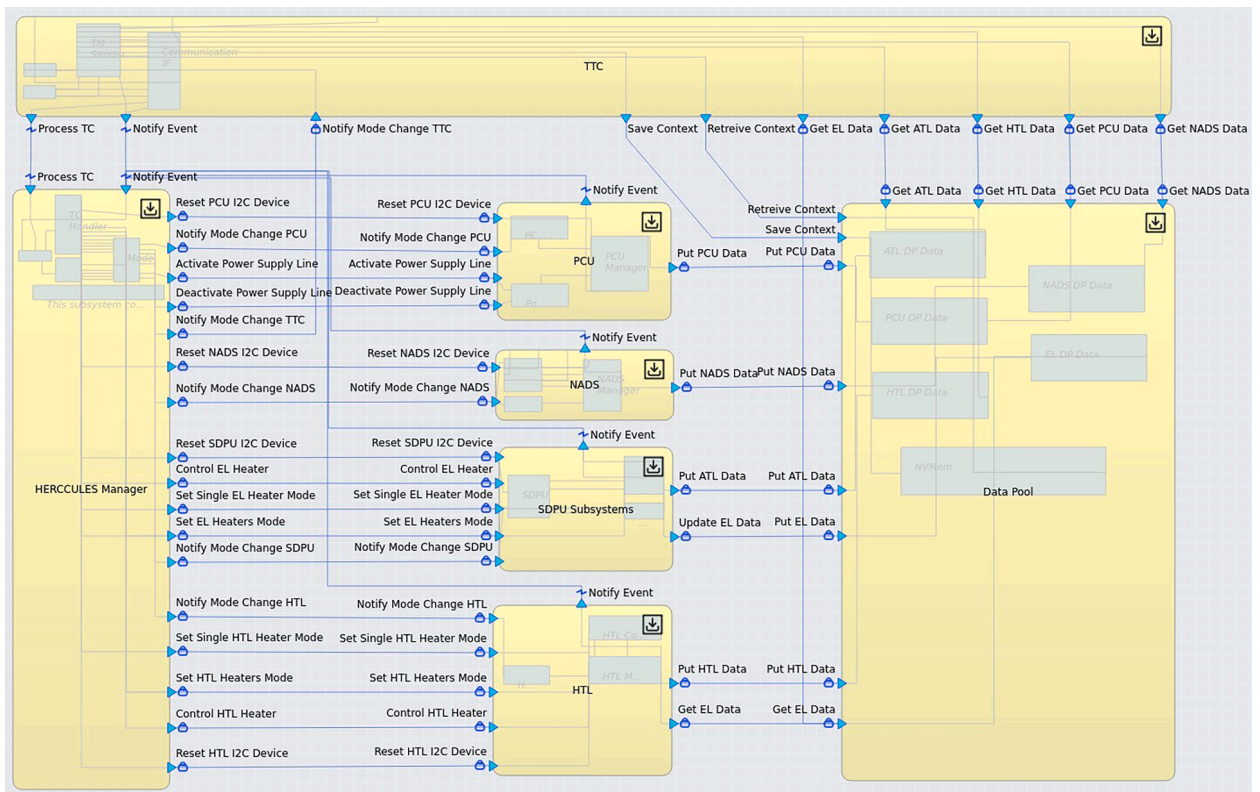

**Fig. 10.** HERCCULES OBSW implemented in TASTE.

Low-level layers were autogenerated by TASTE and implemented real-time abstractions independent of the execution platform.

*5.4.2. Event-driven architecture*

The event-driven or event-based architecture was supported by an asynchronous messaging communication where software components sent notifications in the form of events. This architecture typically contains one central "event processor" that receives events sent by "publishers" and processes them for its redirection to the "subscribers". In general computing applications, the publishers, subscribers, and event topics can be created at runtime, following a dynamic topology. This is inconceivable for strict RTES since it complicates the RTA, as the response times and inter-task communications may vary at runtime. Conversely, a static configuration affords more robust compile-time correctness assurances. Hence, RTES should use this pattern with a static topology. In the case of space software frameworks, TASTE and F-Prime use a static topology, whereas core Flight System (cFS) opts for a configurable topology as it targets large-scale flight systems. The AADL diagram from Fig. 12-a illustrates the static variant of this architecture. On the left side, two publisher tasks trigger events and send them to the Event_Processor sporadic task through its Event_Queue port. Depending on the received event, new events may be notified to subscribers. In this architecture, subscribers received events through data ports because they were not triggered by the incoming events and they may read the input event multiple times and at their discretion.

*5.4.3. Data pool pattern*

The data pool pattern, also known as Repository or Shared-Data pattern, is used in systems where multiple components access and update a common set of data. This pattern is particularly useful in RTES, where data is shared among multiple tasks. The Data Pool is a software component that gathers the shared data and offers two types of operations, one to read and another to update its value. These operations depend on the tasking model of the programming language. In Ada, protected objects may be used to ensure mutual exclusion, while in C/C++, they can be implemented with mutexes or semaphores. The AADL diagram in Fig. 12-b presents an instantiation of this pattern for a hypothetical thermal control system. On the left, the Measurer updates the temperature, while the Thermal Controller reads its value and updates the heater status based on the performed actuation. On the right, the Telemetry Sender task transmits the latest value from the Temperatures and Heaters to the GS. This pattern is defined in [14,17] and it was successfully applied in missions such as TASEC-Lab [3] and the FLP framework [14]. The HERCCULES OBSW used this pattern to share data across tasks.

## 6. Experimental results and discussion

The V&V process of HERCCULES were compliant with the ECSS-E-ST-10-02C standard for Systems Engineering [9] that includes four methods to carry out V&V: Test, Analysis, Review, and Inspection. The V&V followed an incremental approach at the unit, integration and system level. Integration testing, in turn, was divided into instrument and RTU testing.

*6.1. Static code analysis*

The static code analysis was performed to verify the quality and complexity of hand-written code. They were obtained with the SourceMonitor open-source software. Table 3 presents in the first two columns the metrics considered for this evaluation, the third column presents the actual value (AV), and the fourth column contains the expected value (EV) range for each metric which are based on the limits proposed by the Software metrication handbook from the ECSS (ECSS-Q-HB-80-04A) [18] and the SourceMonitor tool. The last column indicates whether AVs met their EV. In this case, all metrics met ECSS requirements for critical systems (levels A and B) except for the maximum cyclomatic complexity that reached a value of 13. Although it exceeded recommended limit of 10 for critical systems, it was acceptable for non-critical software (levels C and D) that have a maximum value of 15. Such complexity was identified in two functions responsible from the TTC module whose complexity is due to its functionality to detect and recover from errors in the communication with the GS. Overall, these results reveal high source code quality as most values are within the expected ranges for critical systems.

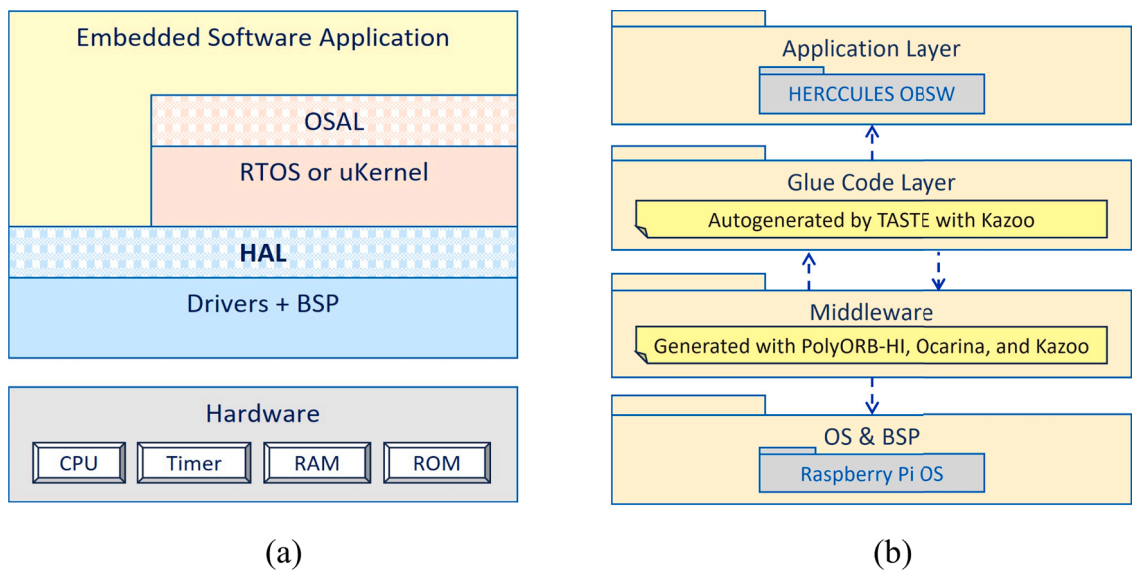

**Fig. 11.** Instances of the layered pattern used in RTES (a) and applications autogenerated by TASTE (b).

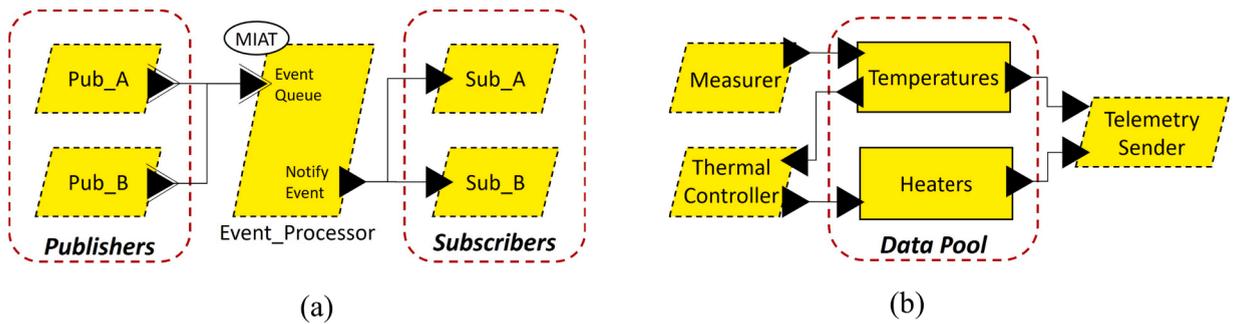

**Fig. 12.** Architecture description in AADL: Event-Driven (a) and Data-Pool/Repository (b).

*6.2. Responsiveness of the system*

The verification of timing requirements for RTES typically involves conducting an RTA to ensure schedulability. Such analysis requires the obtention of the WCET for its tasks and protected objects. However, even with all the OBSW tasks running in the same core, the obtention of the WCET on modern processors (such as the ARM Cortex A4 used in this project) is problematic due to the unpredictability introduced by cache memories or preemptions caused by interrupts [12]. To address this issue, an alternative approach was used by analyzing the tasks' drift obtained from activation log files. The drift of the $n^{th}$ log entry was defined as the difference in its theoretical record time ($TR_n$) and actual record time ($AR_n$). The former was obtained from the $n^{th}$ timestamp recorded on the log file, while the latter was defined as:

$$TR_n = AR_{n=1} + n \cdot P \qquad (1)$$

where $AR_{n=1}$ is the actual record time from the first entry and $P$ is the recording period.

Table 4 summarizes the data collected to assess the responsiveness analysis. The first two columns contain the cyclic tasks and their logs. It was observed that the SDPU Measurer task records data from the ATL and EL subsystems, while the TM Sender records the SC and HK TM logs. The other logs were recorded by dedicated tasks. The third column shows the period at which each log is recorded. The average drift and maximum drift are presented in the third and fourth columns, respectively. Note that this analysis could not be performed on the Manager and TTC tasks since they did not generate any log file. It was obtained that the highest average drift is from the ATL log with $6.1223 \times 10^{-2}$ seconds, and logs with the highest maximum drift were found in the HK TM log with a value of $1.6576 \times 10^{-1}$ seconds. In general, these results suggested that record times of the log entries wer close to their theoretical values. Although the observed maximum drifts values could seem large for typical RTES, they were acceptable given real-time requirements of the HERCCULES OBSW. Additionally, it should be noted that, as imposed by the TASTE runtime, all tasks executed on the same processor core.

**Table 3**
Summary of metrics for HERCCULES OBSW static code analysis.

| Metric | Definition | AV | EV | Success |
|---|---|---|---|---|
| Max. cyclomatic complexity | Number of linearly independent test paths for each subroutine, as defined by McCabe. | 13 | $\leq 10$ | No |
| Avg. cyclomatic complexity | Obtained as the average of code cyclomatic complexities. | 1.78 | $\leq 10$ | Yes |
| Max. function depth | Depth of imbrications of the code. Obtained as the maximum nested block depth level found. At the start of each file, the block level is zero. | 6 | $\leq 6$ | Yes |
| Avg. function depth | Obtained as the average of nested block depths. | 1.43 | $\leq 6$ | Yes |
| Avg. stmts. per method | Average number of executable lines per method (routine in the case of modules). | 4.9 | $\leq 50$ | Yes |
| Avg. methods per class | Average number of public and private methods and procedures per class. | 9.11 | $\leq 20$ | Yes |
| Percentage of comments | Percentage of the number of comments (outside of and within functions). | 17.6 | $\leq 30\%$ | Yes |

**Table 4**
Responsiveness of the system based on activation log's timestamps with time values in seconds.

| Task | Log | Log period | Avg. drift | Max. drift |
| --- | --- | --- | --- | --- |
| IMU Measurer | NADS | 0.01 | $2.3523 \times 10^{-5}$ | $2.6314 \times 10^{-3}$ |
| SDPU Measurer | ATL | 1 | $6.1223 \times 10^{-2}$ | $1.1658 \times 10^{-1}$ |
|  | EL | 1 | $1.0876 \times 10^{-2}$ | $3.3397 \times 10^{-2}$ |
| PCU Manager | PCU | 5 | $1.5106 \times 10^{-5}$ | $1.2302 \times 10^{-4}$ |
| HTL Manager | HTL | 10 | $1.4277 \times 10^{-5}$ | $5.8889 \times 10^{-2}$ |
| TM Sender | SC TM | 1 | $1.7822 \times 10^{-3}$ | $1.5404 \times 10^{-1}$ |
|  | HK TM | 10 | $1.6533 \times 10^{-3}$ | $1.6576 \times 10^{-1}$ |

*6.3. Tests and results at the instrument level*

*6.3.1. Mediator equipment: ADC, multiplexers, and $I^2C$ bus*

The ADC was one of the lowest level components. It was the first component under test because its validity had a direct impact on higher level components. The ADS1115 ADC from the HAL was validated with automated tests that checked the initialization sequence, the configuration of the ADC's operating mode, and the evolution of successive raw readings. Although the correctness of the readings could not be verified as they depended on the devices connected to the ADC (confer the section 6.4), these tests allowed to verify additional aspects related to timing like waiting and transition periods. All automated tests from the ADC controller passed. With regards to timing, the test that verified individual ADC readings took a total of 1043 ms. Considering that the ADC was initially configured with a sampling rate of 1/8 samples per second and that the test included 8 consecutive readings, a total of 1000 ms was expected. The remaining 43 ms were due to the tests overhead. The ADC was connected to the OBC through an $I^2C$ interface and channels were selected through GPIO pins. Hence, These tests allowed to verify the Bus Handlers and Equipment Handler's layers.

*6.3.2. GPS receiver*

HERCCULES was equipped with a Mikroe-1032 GPS Click sensor that carries a U-blox LEA-6S module. It was connected to the OBC through a UART line at 115200 bauds and configured at 5 Hz frequency sending NMEA messages containing position, velocity, and time information. The GPS time was compared with the time obtained via the Network Time Protocol (NTP) which has an accuracy about 128 ms. The latitude and longitude were validated by comparing measured values with the position obtained from another receiver used as the "ground truth". Fig. 13 shows the longitude and latitude that were obtained for 35 seconds. Latitude showed a linear tendency with an average value of 40.437699° that was near the expected value of 40.437700°. Regarding longitude, although the evolution was less noisy, a slightly increasing tendency was observed with an average value of -3.672525° that was close to the reference longitude of -3.672524°. Overall, differences among expected and reference values were small enough for the HERCCULES purposes.

*6.3.3. Additional validated equipment*

First of all, the software that controlled the Adrafruit BNO055 IMU was verified by Inspection through a Graphical User Interface (GUI) autogenerated by TASTE, depicted in Fig. 14. This GUI allowed the NADS operators to inspect the IMU measurements at 1 Hz frequency and send commands for its calibration, restart, and configure parameters such as operating mode, directions, and units of its axes. In general, the results of the inspection proved that the sensor was operational and adequately configured for this mission. On another hand, the MS561101BA03-50 absolute barometer was validated by Test comparing the read measurements with the expected pressure (954 mbar) obtained from the weather monitoring stations of the Madrid city council (Spain). Besides, since the TASEC-Lab mission successfully used these barometers [3], the software that controlled this device was reused, hence, validation by Similarity was applicable to this module. The plot from Fig. 14 depicts the evolution of 100 measurements from the two barometers with the expected value represented by the dashed red line. These results suggested that both sensors had values close to the expected ones with slight deviations that were lower than 0.25 mbar, which corresponded to a percentage error of 0.026%, approximately [3].

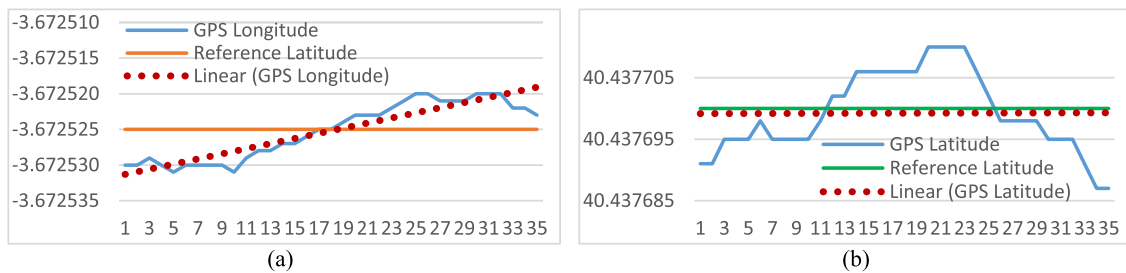

**Fig. 13.** Evolution of the GPS's longitude (a) and latitude (b) compared to the expected values.

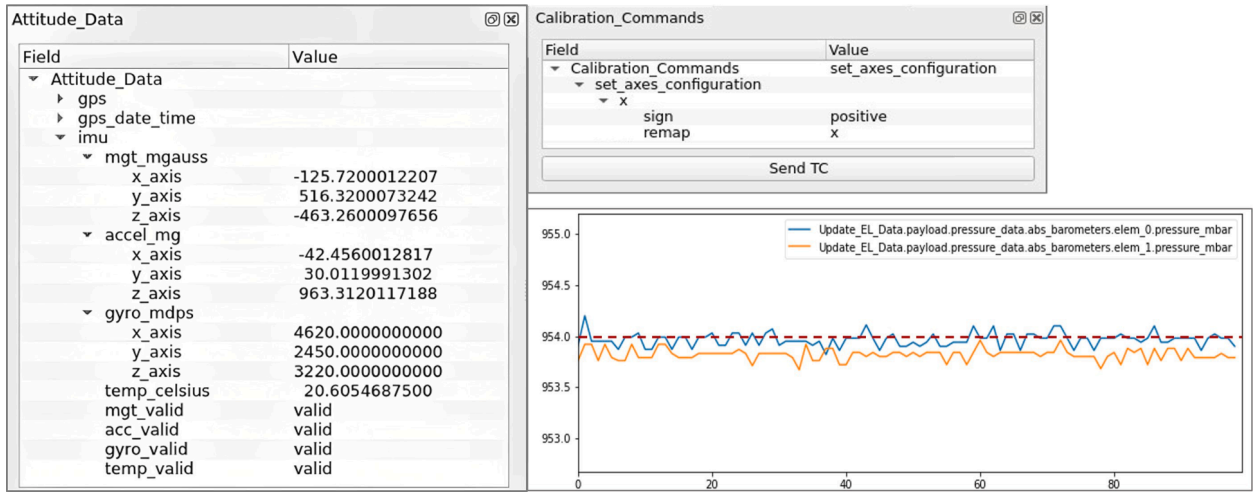

**Fig. 14.** Autogenerated GUI with the control and inspection tabs of the IMU with a sampling frequency of 100 Hz, and a plot of barometers sampled at 1 Hz for 100 seconds.

### 6.4. Verification, validation, and results at the RTU level

#### 6.4.1. TMU testing

The TMU contained the data acquisition circuit for the PT1000 thermistors from HTL. To automate the testing of the TMU software, this RTU was connected to a custom test-bench carrying a series of resistors providing fixed voltages at the input of all twenty-eight channels. Table 5 summarizes the obtained results for multiplexers 0 and 2, other channels (1 and 3) are not presented as they presented a similar behavior. Each row shows the obtained values per multiplexer-channel pair specified in the two first columns. The third column indicates the expected theoretical voltage, while the fourth column shows the average voltage obtained from fifty samples. The fifth column presents the accuracy of these results by indicating the absolute percentage error which was obtained as follows:

$$Error\ (\%) = \frac{|TV - AV|}{TV} \cdot 100\% \quad (2)$$

where $TV$ represents the Theoretical Value and $AV$ the Actual Value, both measured in Volts. The total accuracy of this test was evaluated by calculating the MSE which was obtained as indicated below:

$$MSE = \frac{1}{N} \cdot \sum_{i=0}^{N} (TV - AV)^2 \quad (3)$$

where N is the total number of multiplexer-channel pairs.

The MSE resulted in $9.6475 \times 10^{-5}$. These results showed that the measured voltages were close to the theoretical values, with most errors falling below 1%. However, channel 2 from multiplexer 0 and channel 2 from multiplexer 2 showed higher errors, which were 1.2989% and 5.2242%, respectively. These outliers could be due to noise and other physical factors influencing the measurement.

**Table 5**
TMU results for analog sensors obtained with test benches for multiplexers 0 and 2.

| Multiplexer | Channel | Theoretical Voltage | Actual Voltage | Error (%) |
| --- | --- | --- | --- | --- |
| MUX 0 | 0 | 2.3559 | 2.3553 | 0.0275 |
|  | 1 | 3.8569 | 3.8333 | 0.6117 |
|  | 2 | 0.0445 | 0.0439 | 1.2989 |
|  | 3 | 2.3503 | 2.3449 | 0.2288 |
|  | 4 | 2.3618 | 2.3582 | 0.1529 |
|  | 5 | 2.363 | 2.3616 | 0.0583 |
|  | 6 | 2.3626 | 2.3589 | 0.1574 |
| MUX 2 | 0 | 2.3552 | 2.3534 | 0.0776 |
|  | 1 | 3.8663 | 3.8408 | 0.6608 |
|  | 2 | 0.0396 | 0.0375 | 5.2242 |
|  | 3 | 2.3531 | 2.3472 | 0.2513 |
|  | 4 | 2.373 | 2.3671 | 0.2473 |
|  | 5 | 2.3526 | 2.3475 | 0.2160 |
|  | 6 | 2.3961 | 2.3852 | 0.4554 |

### 6.4.2. SDPU testing

The SDPU implemented the data acquisition pipeline for the ATL, NADS, and EL analog signals. To test the analog lines, the same methodology as that used for the TMU was adopted. A test bench was set up with fixed voltages assigned to each channel, allowing for automated software tests to be conducted. It is noteworthy that only a subset of the lines was utilized for reading the analog signals, and those unused did not need to be tested. Table 6 presents on each row the theoretical voltage, actual voltage, and percentage error for most relevant multiplexer-channel line. The resulting MSE value for these measurements was found to be $0.59471 \times 10^{-6}$. In general, the SDPU testing showed similar results to the TMU testing with most of the errors below 0.1%. However, there were a few readings that showed slightly higher relative errors ranging from 0.13% up to 0.37%, which were small enough for the experiment purposes.

### 6.4.3. PCU testing

The PCU testing differed from the SDPU and TMU since it only involved digital sensors, namely, TC74 thermometer and INA226 voltage and current sensor. However, this test adopted a similar approach, by comparing the read values with their theoretical values. The test cases simulated low-voltage, nominal, and high-voltages situations with supply voltages of 26.0 V, 28.0 V, and 30.0 V, respectively. Regarding the TC74 sensor, its theoretical value was obtained from a previously validated thermometer. Table 7 presents the results from the three test cases in different rows. Each one consisted of four measurements that are presented in the second column, and the theoretical value, actual value, and percentage error are depicted in the last three columns, respectively. In general, the results demonstrated low differences across measurements and, specifically, the voltage (V) and temperature (°C) did not surpass errors of 0.08% and 2%, respectively. On the other hand, current (A) and power (W) exhibited higher errors, with the highest values occurring in the low voltage test case.

### 6.5. System level testing

#### 6.5.1. Software integration testing at ground

The validation of the complete software system was performed by Inspection. This test required the execution of all subsystems managers concurrently and was performed on the proto-flight model of HERCCULES, as depicted in Fig. 15. In these tests, the GUIs autogenerated by TASTE were used to analyze the TM sent by the OBSW and to evaluate the behavior of the subsystems based on commands sent by the operators. These commands included the TCs and special operations to manually inject events and errors such as changes in the pressure, and loss in the communications, among others. These tests allowed to check coarse-grained features such as the operating mode management, performance of subsystems in each mode, and reaction to events and errors. In general, the performance and responsiveness of the system showed similar results as previously discussed. Data was successfully recorded onboard the Raspberry Pi, which allowed to corroborate the expected results after the system execution.

#### 6.5.2. Functional testing at Thermal Vacuum Chamber (TVAC)

So far, all tests described were performed on ground at ambient temperature and air pressure near sea level. However, it is also convenient to validate the system under similar conditions as those experienced during the mission. In this regard, the TVAC test was performed at IDR/UPM facilities, as shown in Fig. 16. During the 8 hours of testing, the pressure was reduced to 11 mbar at a 10 mbar/min rate, which corresponded to the pressure at an altitude of 30 km, which was increased up to the initial pressure at ground of 954 mbar. These changes allowed to verify the air pressure read from the OBSW by comparing them with the TVAC data. Besides, the evolution of pressure during the test emulated the flight profile, which enabled to verify the changes in operating modes automatically triggered by the OBSW. This test also provided the validation of the equipment performance under vacuum conditions and extreme temperatures.

As depicted in Fig. 17, the test was performed twice (a and b). Both figures present the evolution of pressure, power consumption and HTL operating modes during the TVAC testing. As illustrated in Fig. 17-a, the OBSW detected the ascent-1 (red) and float-1 (blue) modes at the expected pressures (900 mbar and 21.5 mbar, respectively). However, after near 4 hours (12500 seconds) of testing, there were two tries to change to float-2, which had to be accessed after a delta time of six hours from the start of float-1. These two failures were due to software errors in the Subsystem component. To verify the updated OBSW efficiently, this test was repeated with a delta time reduced to 20 minutes, as shown in Fig. 17-b. In the updated OBSW, float-2 mode started after 20 minutes (600 seconds) and the shut-down of the experiment during descent was also successfully tested.

**Table 6**
SDPU results for analog sensors obtained with test benches.

| Multiplexer | Channel | Theoretical Voltage | Actual Voltage | Error (%) |
| --- | --- | --- | --- | --- |
| MUX 0 | 0 | 1.0000 | 1.0016 | 0.1600 |
|  | 1 | 1.0000 | 1.0015 | 0.1500 |
|  | 2 | 2.0480 | 2.0466 | 0.0684 |
|  | 3 | 2.0480 | 2.0499 | 0.0928 |
| MUX 1 | 0 | 2.2728 | 2.2739 | 0.0484 |
|  | 1 | 2.2431 | 2.2418 | 0.0580 |
|  | 2 | 3.8292 | 3.8268 | 0.0627 |
|  | 3 | 0.2730 | 0.2740 | 0.3663 |
|  | 4 | 2.3559 | 2.3553 | 0.0255 |
|  | 5 | 3.8367 | 3.8302 | 0.1694 |

**Table 7**
PCU results for analog sensors obtained with test benches.

| Test Case | Measurement | Theoretical Value | Actual Value | Error (%) |
| --- | --- | --- | --- | --- |
| Low Voltage | Voltage (V) | 26.0000 | 26.0175 | 0.0673 |
| | Current (A) | 0.1300 | 0.1206 | 7.2115 |
| | Power (W) | 3.3800 | 3.1375 | 7.1746 |
| | Temperature (°C) | 20.4000 | 20.0000 | 1.9608 |
| Nominal | Voltage (V) | 28.0000 | 27.9963 | 0.0134 |
| | Current (A) | 0.1200 | 0.1126 | 6.1500 |
| | Power (W) | 3.3600 | 3.1531 | 6.1570 |
| | Temperature (°C) | 20.3000 | 20.0000 | 1.4778 |
| High Voltage | Voltage (V) | 30.0000 | 30.0238 | 0.0792 |
| | Current (A) | 0.1000 | 0.1056 | 5.5625 |
| | Power (W) | 3.0000 | 3.1703 | 5.6771 |
| | Temperature (°C) | 20.3000 | 20.0000 | 1.4778 |

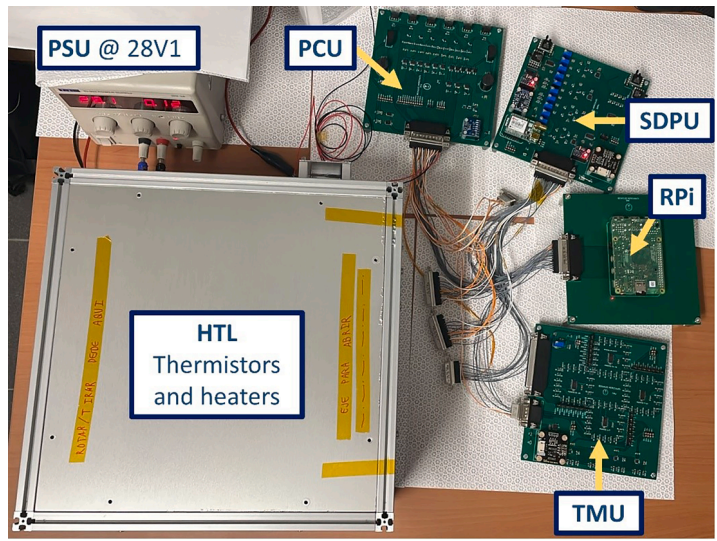

**Fig. 15.** Proto-flight model for HERCCULES validation.

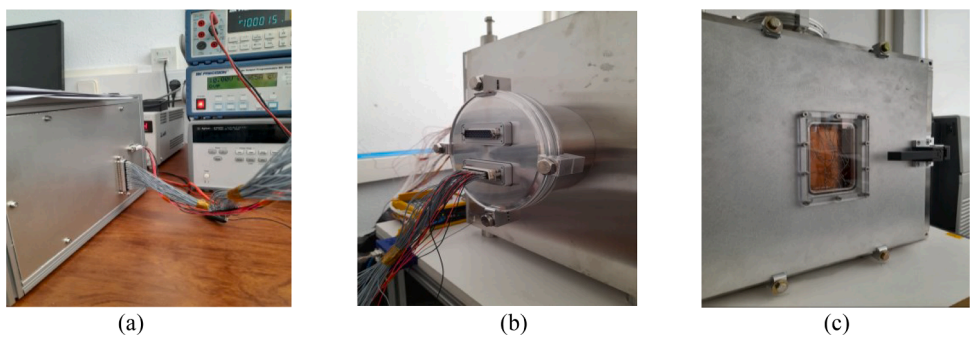

**Fig. 16.** Initial setup with the E-Box inside the TVAC (c) connected via DSUB interfaces (b) to the HTL (a).

During the TVAC testing, additional functionality from the GS was tested. In such test, the GS was connected to the OBC through a cross-over Ethernet cable with a limited bandwidth of 500 kbps, which is the quota available for HERCCULES. The bandwidth consumption was analyzed with the Wireshark tool, resulting in around 1.93 kbps for downlink of TMs and 0.56 kbps for uplink of TCs. These results showed that in terms of bandwidth, HERCCULES was below the maximum limit. In addition, the Ethernet link was

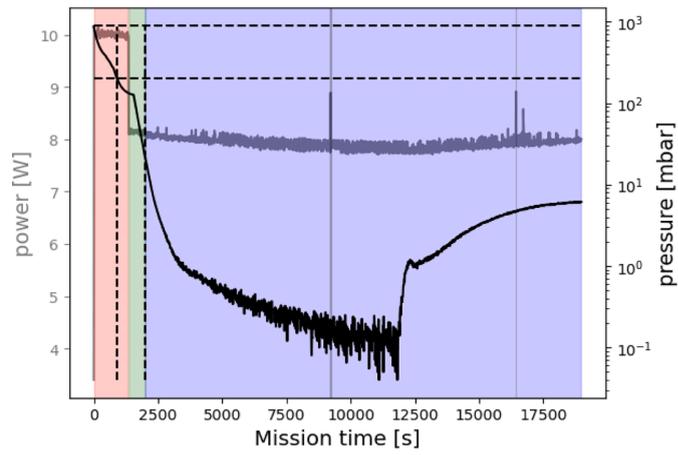
(a) First part (red: ascent-1, green: ascent-2, blue: float-1)

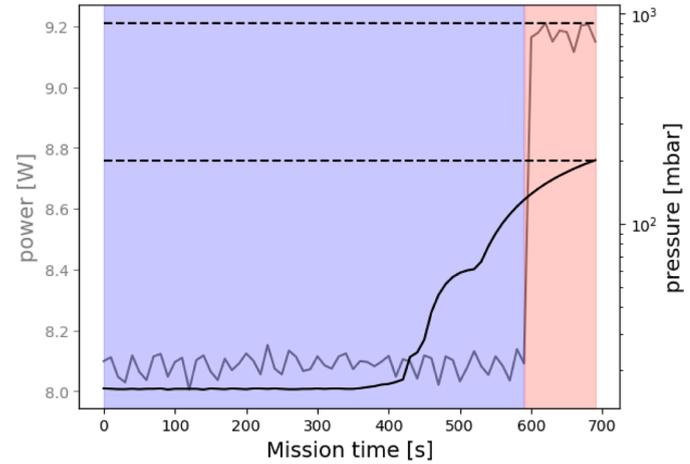
(b) Second part (blue: float-1, red: float-2)

**Fig. 17.** HTL modes evolution when the E-Box was inside the TVAC: first part (a) and second part (b).

connected and disconnected several times to verify the fault tolerance functionality against connection loss. During such tests, OBSW and GS reestablished the connection, and the OBSW updated the control of heaters to autonomous mode per requirements. The Fig. 18 depicts the HK TM represented by the GS during the second part of the TVAC test (Fig. 17-b) and the plot corresponds to air pressure readings from the OBSW.

*6.6. Comparison to related research*

This subsection provides a qualitative and quantitative comparison of the proposed development with related research works. This article includes experiments typically used in weather stations that were developed using methodologies applicable to space systems, such as satellites. Therefore, we considered similar works such as COTS-based monitoring stations on the ground (conventional ground weather stations) and in flight (unmanned aerial vehicles, UAV).

In some stratospheric balloon experiments, the developments often use COTS hardware, such as in [6] with a Microchip PIC microcontroller for A/D conversion and antenna motor control. Like our work, it was developed using a combination of low-cost COTS peripherals such GPS receivers, accelerometers, and magneto-resistive circuits to determine gondola attitude. However, it was not validated following safety-critical methodologies and suffered from failures in the GPS modules. The thermistors employed in that study exhibited a resolution and an accuracy of less than 0.07 °C and 2 °C which were worse than our results of $4.27 \times 10^{-6}$ °C and 0.113 °C, respectively. Another relevant project is the Huygens Cassini stratospheric balloon mission [19] that reached similar altitudes (32 km). In that experiment, the control and data acquisition were performed by a Pentium computer and its OBSW was categorized as a soft RTES that controlled peripherals through A/D boards and conventional buses such as RS-232. Compared to our work, the system had a poorer accuracy for its pressure readings with a percentage error of 1% (ours was 0.25%) and, although it claimed to be a soft RTES with periods of 1-100 Hz, it did not describe the real-time architecture, nor the schedulability assessment of the system.

Similar to the HERCULES OBC, some CubeSat projects used Raspberry Pi boards, such as in [7] to locate an optical ground station beacon. In this case, the OBC was a Raspberry Pi 3B developed without specific development and validation processes for critical systems. Like HERCCULES, the Raspberry Pi executed a Linux based OS in its default configuration and proved to be efficient achieving execution times between 0.298 and 2.010 seconds. Although these time values depend on the executed algorithm, the expected results were obtained with a similar OBC and OS, fulfilling the real-time requirements from the nine tasks (listed in Table 2) with periods ranging from 10 ms (IMU Measurer task) up to 10000 ms (HTL Manager task). Other projects, such as Pathfinder [8], developed communications experiments based on Iridium satellites technology that had to comply with low cost, weight and power requirements. It used a 40 MHz Intel 188 microprocessor as OBC and baremetal OBSW without an OS. The advantage of such approach was the performance and customizability, since low level operations (timers, buses, etc.) were directly accessible by the user. However, the software was tightly coupled to the hardware. The presented work was developed with TASTE that abstracts the underlying platform, making most part of the OBSW portable to other targets.

In most research works related to environmental monitoring, the results are not characterized numerically using metrics such as the MSE or Error (%). Then, it is necessary to compare HERCCULES with COTS-based applications that quantitatively evaluate similar parameters. The Table 8 includes similar COTS-based research works including different types of applications for measuring environmental conditions, such as a high-altitude balloon [20], an UAV [21], or ground-based weather stations [22,23]. Furthermore, a broader perspective of the architectural solutions was obtained by considering other systems related to the space sector such as the control software system for an instrument deployed in the International Space Station (ISS) [24].

Firstly, regarding the V&V methodology, the presented work adopted the Analysis (that includes Similarity and Comparison), Inspection, Review-of-design, and Test as recommended by the ECSS-E-ST-10-02C standard. Considering that Comparison is a subset of Analysis, a large part of the related works only use verification by Comparison [21–23] because they are low criticality systems, only one related work uses Analysis which is proper for medium criticality systems [20], and in those critical developments used in space applications for onboard instruments [24] the Analysis, Review-of-design and Test methods are adopted using complementary tools, such as Codacy. Based on these works, it can be concluded that the V&V methodology used in HERCCULES is more oriented to critical space systems, and although the system is not critical as such, a validation was carried out with high quality considerations using the SourceMonitor static code analyzer and additional tools that supported automated testing such as g-test.

Secondly, regarding the system architecture, most systems are developed with COTS devices and technology. Specifically, some of the works use a Raspberry Pi as the central OBC because of its powerful features and the support of Linux-based OS [22]. Specifically, Linux-based OS is convenient for applications that require concurrent measurement of environmental parameters as it provides scheduling policies for RTES. In space projects with higher criticality levels, more specific COTS processors are used such as the Intel Atom and Xilinx Zynq XC7Z030 coprocessor for the data acquisition system of an onboard instrument [24]. In the rest of the analyzed works, boards based on Atmel microcontrollers, including Arduino, are used as central OBCs such as in a high-altitude balloon as technology evaluator for CubeSat missions [20], environmental monitoring using UAVs [21], and a weather station [23]. In the presented work, the selected OBC aboard a high-altitude balloon was configured with a Linux OS making use of the POSIX profile for RTES.

Thirdly, regarding the obtained results, it is difficult to compare this matter because metrics not only depend on the performance or quality of the OBSW, but also on the entire system including hardware architecture. Concerning source code quality, the Mini-EUSO experiment [24] specifies the highest level (Grade A) using the Codacy proprietary tool. HERCCULES opted for open-source software to obtain such metrics and, though an overall grade was not provided, a high-quality level was deducted based on ESA recommendations. The obtained MSE for TMU had a value of 0.1 °C, which is better than in an environmental monitoring station with RMSE values of 2.15

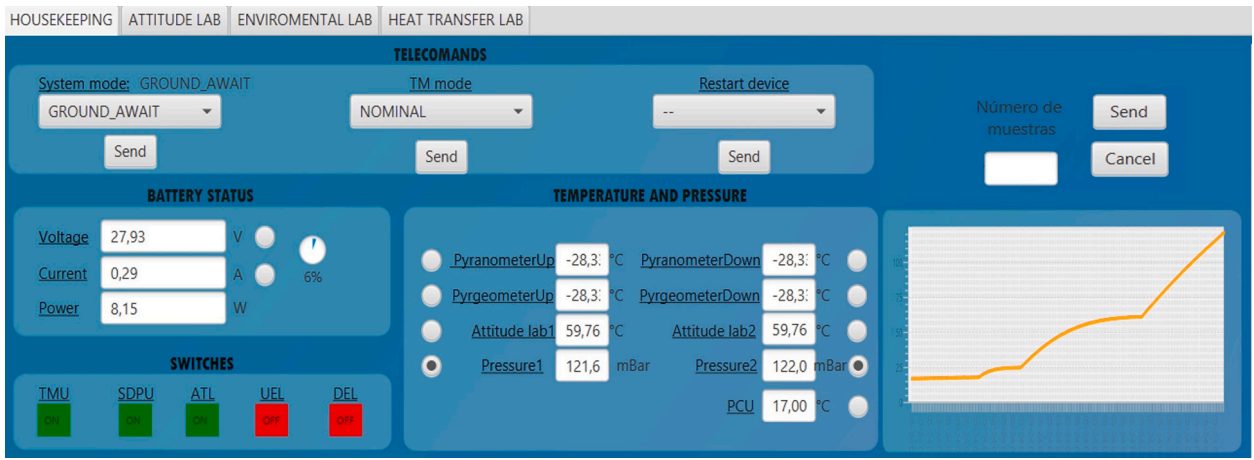

**Fig. 18.** Housekeeping tab from the customized Ground Station.

°C (MSE of 4.62 °C) for temperature, 1.31 °C (MSE of 1.72 °C) for atmospheric temperature-1, and 0.91 °C (MSE of 0.83 °C) for atmospheric temperature-2 [22]. The obtained MSE also surpassed the results of a weather station that obtained an MSE of 0.97 °C for air temperature [23]. In addition, a mean absolute error (MAE) of 1.4 °C for temperatures within 11-20 km is obtained for a high-altitude balloon application [20], which is above the MAE of 0.2 °C for the TMU. The remaining data is difficult to compare with other experiments; however, a similar behavior is obtained with percentage errors below 6% such as for a UAV sensing system [21]. In general, considering the level of criticality and the HERCCULES requirements, the results were obtained within the specified limits with relatively low errors, and with a negligible influence on the performance of the experiment.

Finally, regarding the development methodology, most of the related works are focused on a system level with no importance relative to the validation of software development, so it is not described in [20–23]. [24] includes the only development that aims at explaining in detail the software architecture due to the criticality of the system, following an Agile paradigm making use of the continuous integration approach. Although the presented work is a non-critical system, it followed the V-methodology recommended by ESA for the development of space systems [17].

*6.7. Limitations*

First of all, the schedulability analysis of the system was limited to a drift analysis that was suitable for a soft RTES as HERCCULES, but it could not be adequate for safety critical systems where analytical demonstrations such as the RTA are mandatory. Besides, tools that estimated the WCET of tasks and shared resources could not be afforded. Secondly, compared to other subsystems such as the TMU or SDPU, the V&V of the GPS receiver and IMU was restricted to a rudimentary approach as there was a no reliable "ground truth". Regarding the TASTE toolchain, despite its advantages as automatic code generation, the tool is relatively new and, thus, some issues were detected while developing HERCCULES. In addition, the quantitative comparison with other studies is difficult as many of them lacked statistical metrics such as MSE or percentage error, and if they do, most of the measurements are not comparable with HERCCULES results. Finally, the qualitative comparison is also complicated since related works barely describe their software design and methodologies.

*6.8. Future works*

The autogenerated code is found to be overloaded with several wrappers and indirection layers. Although it works well on HERCCULES, it needs to be optimized for platforms with limited computing resources, which further highlights the merit of its use in this study. The TASTE tool supports real-time abstractions, but its current version does not ease the calculation of the execution times of tasks and resources. Therefore, the automatic calculation of such values would help to estimate the WCET and in turn automate the RTA process. Another interesting research line is related to the automatic generation of GS software taking the TM and TC defined in ASN.1, which can be improved by working on the autogeneration of web servers, generic GUIs, databases, and also in the automatic deployment of such services with Docker containers. These works would be valuable for the development of the future UPMSat-3 microsatellite in which the authors are participating.

**7. Conclusions**

This article presents the methodology and development of the OBSW system for the COTS-based HERCCULES stratospheric balloon. The OBSW is developed with the TASTE toolchain following the MBD and CBD methodologies. The OBSW is verified at different granularity levels and applying multiple V&V methods. Compared to similar works, HERCCULES is designed through several review processes and results are verified using quantitative metrics such as MSE or percentage error. The static analysis reveals high-quality source code per the recommendations of ECSS-Q-HB-80-04A for critical software. The tasks' responsiveness is assessed by drift analysis, demonstrating the system predictability with 0.1 seconds as the highest drift in the SDPU Manager task. At the RTU level, obtained results are within specified limits with a percentage error below 6% in the temperature readings of the TMU and SDPU. At the system level, the overall functionality is verified, including concurrent mode management and events and errors handling. Compared to similar research works, the OBSW architecture design considers real-time aspects; for reference only, the Linux OS is used leveraging the POSIX profile for RTES and the OBSW is designed applying three design patterns suitable for RTES.

**Table 8**
Comparison of the system proposed in the present work to related research.

| Ref. | System architecture | V&V methodology | Results |
| --- | --- | --- | --- |
| Present work | Stratospheric balloon application. Equipment: COTS (radiometers, thermistors, barometers…) with Raspberry Pi 4B as OBC and communications with a remote GS. | By Analysis, Comparison, Similarity. By Review-of-design, code metrics. By Inspection, operators via the GS. By Test, automated at different levels. | $MSE_{imu\_volt} = 9.647 \times 10^{-5}$ $V$; $MSE_{imu\_deg} = 1.127 \times 10^{-1}\ °C$; $\max(\%\ Error_{pcu\_v}) = 0.08\%$; $Press\ AE_{on\_gnd} = 0.25\ mbar$ |
| [20] | High-altitude balloon for CubeSat missions. Equipment: Arduino Uno as OBC, DHT22 for temperature, and BME280 for pressure. | By Comparison, the BME280 pressure compared to readings from a GPS. By Analysis the temperature is checked based on a standard atmosphere model. | $AltitudeAE_{at16km} = 897m$; $AltitudeAE_{ongnd} = 88m$; $TempAE_{at11-20km} = 1.4°C$ |
| [24] | Mini-EUSO instrument of the ISS. Equipment: Intel Atom E3815 as OBC, thermistors, photodiodes, and the Zynq XC7Z030 system-on-chip. | By Analysis, instrument in simulations. By Review, code quality with Codacy tool. By Test with automated and manual tests. | $CodeQuality = GradeA$; $Availability > 95\%$ |
| [21] | UAV sensing system for particulate matter (PM). Equipment: Atmega 2560 as OBC connected to a GPS. The PMS5003RD sensor measures PM2.5 and PM10 particulates. | By Analysis/Comparison with an air-quality monitoring station. A Ground Station was used to present and record received data for further Inspection and Analysis. | $Error_{PM2.5} = 6.2\%$; $Error_{PM10} = 6.6\%$; $MRE_{PM2.5} = 19.2\%$; $MRE_{PM10} = 27\%$ |
| [22] | Environment monitoring station for atmospheric and surface temperature. Equipment: Raspberry Pi as OBC and COTS sensors (DS18B20, DHT11). | By Analysis/Comparison, measured values from the Sentinel-3 satellite and a weather station located nearby. The root MSE (RMSE) is obtained for all measured data. | $RMSE_{temperature} = 2.15\ °C$; $RMSE_{atmtemp1} = 1.31°C$; $RMSE_{atmtemp2} = 0.91°C$ |
| [23] | COTS-based weather station for natural disaster monitoring. Equipment: Arduino Mega 2560 as controller, anemometer, thermometers, barometers, and humidity sensors. | By Analysis/Comparison, measured values compared to data from reference weather station located 3m apart the system. The data analysis involves various performance metrics including the MSE and RMSE. | $MSE_{airpress} = 0.2815mbar$; $MSE_{airtemp} = 0.9686°C$; $MSE_{humidity} = 14.689\%$; $MSE_{windspeed} = 0.6404m/s$ |

In summary, this study presents a methodology and development of a non-critical OBSW system that adheres to guidelines for space systems and improves development time, correctness, and reliability. Specifically, the MBD paradigm proves to be a valuable option due to its code generation capabilities, which reduce costs, errors, and development effort. The CBD paradigm and design patterns promote composability, verifiability, and reusability of software components. In addition, the incremental validation philosophy allows for early and rapid development, and while the V-model is commonly used for safety-critical software, this article demonstrates its effectiveness on non-critical systems like HERCCULES.

**Code availability**

The source code as well as the TASTE and ASN.1 models for the OBSW and GS are distributed as open-source software in GitLab repositories [25] under the GNU General Public License v3.0.

**CRediT authorship contribution statement**

**Ángel-Grover Pérez-Muñoz:** Investigation, Software, Methodology, Resources, Conceptualization, Formal analysis, Validation, Visualization, Writing – original draft, Writing – review & editing. **Jose-Carlos Gamazo-Real:** Investigation, Supervision, Methodology, Funding acquisition, Conceptualization, Formal analysis, Visualization, Writing – original draft, Writing – review & editing. **David González-Bárcena:** Investigation, Project administration, Writing – original draft. **Juan Zamorano:** Investigation, Project administration.

**Declaration of Competing Interest**

The authors declare the following financial interests/personal relationships which may be considered as potential competing interests:

Jose-Carlos Gamazo-Real reports article publishing charges was provided by the project OAPES-CM "Operación Avanzada de Pequeños Satélites" (Ref.: Y2020/NMT-6427) of the IDR/UPM and CRUE Spanish Universities - CSIC Alliance. Jose-Carlos Gamazo-Real reports a relationship with Polytechnic University of Madrid (Universidad Politécnica de Madrid, Spain) that includes: employment.

**Data availability**

CODE AVAILABILITY: Source code, and TASTE and ASN.1 models are available GitLab=> https://gitlab.com/AngelPerezM/herccules/-/tree/main/src AND https://gitlab.com/herccules/.

**Acknowledgement**

This work was mainly supported by the IDR/UPM under the project OAPES-CM "Operación Avanzada de Pequeños Satélites" (Ref.: Y2020/NMT-6427). In addition, the work was developed within the projects AURORA (101004291) and PRESECREL (PID2021-124502OB-C43). The authors also acknowledge the financial support of the European H2020 program, Ministerio de Ciencia e Innovación, and Comunidad de Madrid Proyectos Sinérgicos from the I+D plan (Spain).

**References**

[1] Abe T, Imamura T, Izutsu N, Yajima N. Scientific ballooning. New York, NY: Springer New York; 2009. https://doi.org/10.1007/978-0-387-09727-5.
[2] Gonzalo J, López D, Domínguez D, García A, Escapa A. On the capabilities and limitations of high altitude pseudo-satellites. Prog Aerosp Sci 2018;98:37–56. https://doi.org/10.1016/j.paerosci.2018.03.006.
[3] González-Bárcena D, Peinado-Pérez L, Fernández-Soler A, Pérez-Muñoz ÁG, Álvarez-Romero JM, Ayape F, Martín J, Bermejo-Ballesteros J, Porras-Hermoso Á, Alfonso-Corcuera D, Marín-Coca S, Soto-Aranaz M, Boado-Cuartero B, Garcia-Romero R, Pindado S, Pérez-Álvarez J, Zamorano J, Torralbo I, Piqueras J, Pérez-Grande I, Sanz-Andrés. TASEC-Lab: A COTS-based CubeSat-like university experiment for characterizing the convective heat transfer in stratospheric balloon missions. Acta Astronaut 2022;196:244–58. https://doi.org/10.1016/j.actaastro.2022.04.028.
[4] Shen Y, Wang H, Blaabjerg F, Zhao H, Long T. Thermal modeling and design optimization of PCB Vias and Pads. IEEE Trans Power Electron 2020;35:882–900. https://doi.org/10.1109/TPEL.2019.2915029.
[5] Perrotin M, Conquet E, Delange J, Schiele A, Tsiodras T. TASTE: a real-time software engineering tool-chain overview, status, and future. In: Ober Iulian, Ober Ileana, editors. SDL 2011: Integrating System and Software Modeling. Springer; 2011. p. 26–37. https://doi.org/10.1007/978-3-642-25264-8_4.
[6] Johansson G, Selinder J, Hyyppa K. DE-Link, an antenna pointing system for stratospheric balloons. In: 2005 12th IEEE International Conference on Electronics, Circuits and Systems. IEEE; 2005. p. 1–4. https://doi.org/10.1109/ICECS.2005.4633446.

[7] Medina I, Hernández-Gómez JJ, Torres-San Miguel CR, Santiago L, Couder-Castañeda C. Prototype of a computer vision-based cubesat detection system for laser communications. Int J Aeronautic Space Sci 2021;22:717–25. https://doi.org/10.1007/s42405-020-00320-4.
[8] Said MA, Stuchlik D, Corbin B, Smolinski M, Abresch B, Shreves C, Stancil R, Cathey H, Cannon S. Overview of the development of the pathfinder ultra-long duration balloon system. Adv Space Res 2004;33:1627–32. https://doi.org/10.1016/j.asr.2003.10.031.
[9] European Cooperation for Space Standardization. ECSS-E-ST-10-02C – space engineering verification 2009.
[10] Wang B, Wang X, Wang N, Javaheri Z, Moghadamnejad N, Abedi M. Machine learning optimization model for reducing the electricity loads in residential energy forecasting. Sustain Comput: Inf Syst 2023;38:100876. https://doi.org/10.1016/j.suscom.2023.100876.
[11] BEXUS members. BEXUS user manual 2021. https://rexusbexus.net/wp-content/uploads/2021/06/BX_REF_BEXUS_User-Manual_v8_03Jun21.pdf (accessed June 15, 2023).
[12] Burns A, Wellings A. Analysable real-time systems: programmed in Ada. 4th ed. Massachusetts: Adisson Wesley; 2016.
[13] Gaber K, El_Mashade MB, Aziz GAA. Hardware-in-the-loop real-time validation of micro-satellite attitude control. Comput Electric Eng 2020;85:106679. https://doi.org/10.1016/j.compeleceng.2020.106679.
[14] Eickhoff J. The FLP microsatellite platform. Cham: Springer International Publishing; 2016. https://doi.org/10.1007/978-3-319-23503-5.
[15] Zamorano Flores JR, Garrido Balaguer J, Cubas Cano J, Alonso Muñoz AA, de la Puente Alfaro JA. The design and implementation of the UPMSAT-2 attitude control system, 50. Elsevier; 2017.
[16] Burns A, Dobbing B, Vardanega T. Guide for the use of the Ada ravenscar profile in high integrity systems. Ada Lett 2004;XXIV:1–74. https://doi.org/10.1145/997119.997120.
[17] Eickhoff J. Onboard computers, onboard software and satellite operations. Berlin, Heidelberg: Springer Berlin Heidelberg; 2012. https://doi.org/10.1007/978-3-642-25170-2.
[18] European Cooperation for Space Standardization. ECSS-Q-HB-80-04A – software metrication handbook 2011.
[19] Bettanini C, Fulchignoni M, Angrilli F, Lion Stoppato PF, Antonello M, Bastianello S, Bianchini G, Colombatti G, Ferri F, Flamini E, Gaborit V, Aboudan A. Sicily 2002 balloon campaign: a test of the HASI instrument. Adv Space Res 2004;33:1806–11. https://doi.org/10.1016/j.asr.2003.07.047.
[20] Lay KS, Li L, Okutsu M. High altitude balloon testing of Arduino and environmental sensors for CubeSat prototype. HardwareX 2022;12:e00329. https://doi.org/10.1016/j.ohx.2022.e00329.
[21] Wang T, Han W, Zhang M, Yao X, Zhang L, Peng X, Li C, Dan X. Unmanned aerial vehicle-borne sensor system for atmosphere-particulate-matter measurements: design and experiments. Sensors 2019;20:57. https://doi.org/10.3390/s20010057.
[22] Kpobi EK, Foli BAK, Agyekum KA, Wiafe G. Development of a raspberry Pi–based remote station prototype for coastal environment monitoring. Remote Sens Earth Syst Sci 2022;5:14–25. https://doi.org/10.1007/s41976-021-00053-2.
[23] Bernardes GFLR, Ishibashi R, Ivo AAS, Rosset V, Kimura BYL. Prototyping low-cost automatic weather stations for natural disaster monitoring. Digit Commun Netw 2022. https://doi.org/10.1016/j.dcan.2022.05.002.
[24] Capel F, Belov A, Cambiè G, Casolino M, Fornaro C, Klimov P, Marcelli L, Piotrowski L, Turriziani S. Mini-EUSO data acquisition and control software. J Astron Telesc Instrum Syst 2019;5:1. https://doi.org/10.1117/1.JATIS.5.4.044009.
[25] Pérez-Muñoz ÁG, Morán A, Garrido-Sola J. GiLab project for the HERCCULES ground system and utility components. https://gitlab.com/herccules/ (accessed April 28, 2023).

**Ángel-Grover Pérez-Muñoz** is an Assistant Professor at the Escuela Técnica Superior de Ingenieros Informáticos of the UPM, and a researcher of the STRAST/UPM group (Madrid, Spain). He is currently pursuing a Ph.D. in Computer Science from the UPM. His main research fields include real-time embedded systems, software for aerospace systems, and artificial intelligence safety.

**Jose-Carlos Gamazo-Real** is an Assistant Professor at the Escuela Técnica Superior de Ingeniería de Sistemas Informáticos of the UPM, Department of Architecture and Technology, and a researcher at the IDR/UPM (Madrid, Spain). He completed a Ph.D. in Information Technologies and Telecommunications Engineering from the Universidad of Valladolid (Valladolid, Spain). His main research interests include electronics and artificial intelligence for aerospace systems, electric motors, and Internet of Things.

**David González-Bárcena** is an Assistant Professor at the Escuela Técnica Superior de Ingeniería Aeronáutica y del Espacio of the UPM, and a researcher at the IDR/UPM (Madrid, Spain). He completed a Ph.D. in Aerospace Engineering from the UPM. His main research interests include thermodynamics, space thermal control engineering, and artificial intelligence.

**Juan Zamorano** is a retired Associate Professor at the UPM (Madrid, Spain). He was a researcher of the STRAST/UPM group at and has collaborated in a number of national and international research projects. He has authored or co-authored over a hundred of technical papers and reports, most of them in the fields of Ada, and real-time systems.